\documentclass[12pt,preprint]{aastex}
\usepackage{psfig,epsfig}

\begin{document}


\title{Mergers of neutron star black hole binaries with small mass ratios:
  nucleosynthesis, gamma-ray bursts and electromagnetic transients}
\author{S. Rosswog$^{1}$\\
         $^{1}$ School of Engineering and Science, International University
         Bremen, Germany
         }


\def\Mesz{M\'esz\'aros~}
\def\Pacz{Paczy\'nski~}
\def\Kluz{Klu\'zniak~}
\def\p{$e^\pm \;$}
\def\msun{M$_{\odot}$}
\def\Msun{M$_{\odot}$ }
\def\be{\begin{equation}}
\def\ee{\end{equation}}
\def\bi{\begin{itemize}}
\def\ei{\end{itemize}}
\def\bea{\begin{eqnarray}}
\def\eea{\end{eqnarray}}
\def\gcc{gcm$^{-3}$}
\def\edo{\end{document}}

\begin{abstract}
I discuss simulations of the coalescence of black hole neutron star binary
systems with black hole masses between 14 and 20 \msun. The calculations use
a three-dimensional smoothed particle hydrodynamics code, a
temperature-dependent, nuclear equation of state and a multi-flavor neutrino
scheme. General relativistic effects are mimicked using the \Pacz-Wiita
pseudo-potential and gravitational radiation reaction forces.\\ 
Opposite to previous, purely Newtonian calculations, in none of the explored
cases episodic mass transfer occurs. The neutron star is always completely
disrupted after most of its mass has been transferred directly
into the hole. For black hole masses between 14 and 16 \Msun an accretion disk
forms, 
large parts of it, however, are inside the last stable orbit and therefore
falling with large radial velocities into the hole. These disks are (opposite
to the neutron star merger case) thin and -apart from a spiral shock-
essentially cold. For higher mass black holes ($M_{\rm BH} \ge 18$ \msun)
almost the complete neutron star disappears in the hole without forming an
accretion disk. In these cases the surviving material is spun up by tidal
torques and ejected as a half-ring of neutron-rich matter. None of the
investigated systems is a promising GRB central engine. We find between 0.01
and 0.2 \Msun of the neutron star to be dynamically ejected. Like in a type Ia
supernova, the radioactive decay of this material will power a light curve
with a peak luminosity of a few times $10^{44}$ erg/s. The maximum will be
reached about three days after the coalescence and will be mainly visible in
the optical/near infrared band. The coalescence itself may produce a precursor
pulse with a thermal spectrum of $\sim 10$ ms duration.

\end{abstract}


\maketitle

\section{Introduction}

Neutron star binary systems have been recognized as potential central engines of
gamma-ray bursts (GRBs) already two decades ago, they have been mentioned in
\Pacz (1986) and Goodman et al. (1987) and discussed in more detail by
Eichler et al. (1989) (for a more complete bibliography we refer to existing reviews,
e.g. Meszaros 2002 or Piran 2005). As a variant of the neutron star binary
case \Pacz (1991) discussed systems containing a neutron star (NS) and a
stellar mass black hole (BH) as a possible GRB engine. These days, such
compact binary systems are considered  the 
'standard model' for the subclass of short gamma-ray bursts, that last
typically for about 0.3 s (Kouveliotou et al. 1993). Most recently, on 9 May
2005 the first ever X-ray afterglow for a short ($\sim$ 30 ms), hard GRB
(GRB050509b) has been detected (Bloom et al. 2005). Its tentative association
with a nearby giant elliptical galaxy has been interpreted as an indication
for a compact binary merger origin of this burst (Bloom et al. 2005, Lee et al. 2005b). 
While NSBH binaries are
usually just considered to be a minor variation on the topic of double neutron
star merger (DNS), it has been pointed out recently (Rosswog et al. 2004) that
it is not obvious, that such a coalescence will automatically produce a hot and
massive accretion disk around the hole. Therefore, its role for gamma-ray
bursts needs further investigations.\\ 
During the disruption process tidal torques are expected to eject material
into highly eccentric, possibly unbound orbits. This debris is extremely
neutron rich, $Y_e \sim 0.1$, and therefore (if ejected at appropriate rates) 
holds the promise to be one of the still much debated sources of r-process
elements (Lattimer and Schramm 1974 and 1976).\\  
Moreover, NSBH systems are generally considered promising sources for
ground-based gravitational wave detectors such as LIGO \citep{abramovici92},
GEO600 (Luck et al. 2001), VIRGO (Caron et al. 1997) and TAMA (Tagoshi et
al. 2001).\\ 
It is worth pointing out in this context that there is a controversy
about the rates at which NSBH mergers do occur. Bethe and Brown (1998)
argued that NSBH should merge about an order of magnitude more frequently than
DNS, while a recent study by Pfahl et al. (2005) comes to the conclusion that
the number of NSBH systems in the Galaxy should be below 1 \% of the number of
double neutron star systems. To date 8 DNS have been observed (Stairs 2004),
while not a single NSBH binary has been discovered yet.\\ 
Neutron star black hole merger simulations have been performed by several
groups. Most of the simulations use Newtonian or pseudo-Newtonian gravity.
Janka et al. (1999) used a grid based hydrodynamics code together with
a nuclear equation of state (EOS) and a neutrino leakage scheme to explore the
role of these mergers in GRB context. Lee (2000, 2001) and Lee and Kluzniak
(1999a,b) used smoothed particle hydrodynamics with polytropic equations of
state to explore the sensitivity of the results to the adiabatic exponent of
the EOS. Lee and Ramirez-Ruiz (2002) have analyzed the flow pattern within an
accretion disk around a BH and, in recent papers (Lee and Ramirez-Ruiz 2004,
Lee et al. 2005), they included detailed microphysics in their
simulations. Setiawan et al. (2004) constructed disks around stellar mass
black holes and followed their evolution including explicit viscosity, Rosswog
et al. (2004) investigated the dynamics of the accretion process. It is worth
pointing out that both Setiawan et al. (2004) and Lee et al. (2005)
implicitly assume that interesting disks form whose properties they
subsequently investigate whereas the focus of this investigation is the question
whether/what kind of a disk forms during the merger.\\ 
Recently, there has been progress in the relativistic treatment black hole
neutron star binaries. Taniguchi et al. (2005), for example, have been
constructing quasi-equilibrium black hole neutron star binaries in general
relativity, and further efforts with approximate relativistic treatments seem
to be underway (Rasio et al. 2005).\\  
Stellar mass black holes formed in core collapse supernovae are thought to be
born with masses ranging from about 3 to 20 \Msun (Fryer and Kalogera
2001). Numerical simulations have so far only explored black hole masses up to
14 \Msun  (Janka et al. 1999 and Lee and
Kluzniak 1999a,b and Lee 2000, 2001 used black holes up to 10 \msun,  Rosswog
et al. 2004 explored masses up to 14 \msun). In this paper we will focus on
black holes with masses ranging from 14 to 20 \msun. The 14 \Msun case has
been explored previously (Rosswog et al. 2004) using the same microphysics but
a purely Newtonian BH-potential and may therefore serve to gauge the effect of
the \Pacz-Wiita pseudo-potential.\\
A simple estimate for the radius where a star around a BH is disrupted is the
tidal radius, R$_{\rm tid}= \left(\frac{{M}_{BH}}{M_{NS}}\right)^{1/3} R_{\rm
  NS}$. As it grows slower with the black hole mass ($\propto$ M$_{\rm BH}^{1/3}$)
than the gravitational radius ($\propto$ M$_{\rm BH}$), it is expected that
higher mass BHs will have more difficulties building up massive disks. 
Therefore, the lower end of the black hole mass distribution is most promising for
the launch of a GRB. Unfortunately, it is very difficult to find suitable
approximations for these cases, as here the space-time is far from a static
Schwarzschild/Kerr solution and therefore neither the use of pseudo-potentials
nor solving the hydrodynamics equations in a fixed background metric are
admissible. For these cases fully dynamical general relativistic calculations are
needed.\\
In this paper we will explore the accretion dynamics and
the observational signatures of the high mass end of these binary systems. 

\section{Simulations}

We use a 3D smoothed particle hydrodynamics code that has been developed to
simulate compact objects. Most of the code features have been described elsewhere
(Rosswog et al. 2000, Rosswog and Davies 2002, Rosswog and Liebend\"orfer
2003) and will only be briefly mentioned for the sake of completeness.
For the simulations described here we have changed to an integration scheme
with individual time steps and also the treatment of the black hole is
different from the previous implementation.\\
In our implementation we have taken particular care to avoid artifacts from
the use of artificial viscosity. The quantities $\alpha$ and
$\beta$ that are usually used as fixed parameters in the artificial viscosity
tensor (Monaghan and Gingold 1983), are made time dependent and evolved by
solving an additional differential equation (Morris and Monaghan 1997). In the
absence of shocks $\alpha= {\alpha}^{\ast}=0.1$ (we always use
$\beta(t)=2\cdot \alpha(t)$), to be 
compared with the 'standard values' of $\alpha= 1$ and $\beta=2$. In strong
shocks our $\alpha$ is allowed to rise up to values of 2 in order
to avoid post-shock oscillations (Rosswog et al. 2000). In addition, the
Balsara prescription (Balsara 1995) 
is used to avoid spurious forces in pure shear flows. Thus, artificial
viscosity is essentially absent unless a shock is detected. These measures have
proved very effective, suppressing unwanted effects of artificial viscosity by
orders of magnitude (Rosswog et al. 2000; Rosswog and Davies 2002).\\
We use a temperature-dependent nuclear equation of state that is based on the
tables provided by Shen et al. (1998a, 1998b) and that has been smoothly
extended to the low-density regime (Rosswog and Davies 2002). It covers a
density range in $\log(\rho)$ from 0.5 to 15.4, temperatures from 0 to 100 MeV
and electron fractions, $Y_e$, from 0 to 0.5. The nucleons are treated in the
framework of temperature-dependent relativistic mean field theory. 
At densities below about 1/3 of nuclear matter density nuclei will be present
in the plasma. Its composition is determined for fixed $\rho$, $T$ and $Y_e$
from an equilibrium of nucleons, alpha particles and an average heavy
nucleus. Thus, energy release due to nuclear transmutations is accounted for
in a simple way. Contributions of photons and electrons/positrons of arbitrary
degree of relativity and degeneracy are added to the thermodynamic 
quantities. Details can be found in Rosswog and Davies 2002.\\
Nowhere (apart from the initial neutron stars) $\beta$-equilibrium is
assumed.
Local changes in the electron fraction and the thermal energy content of
matter due to neutrinos are accounted for with a multi-flavor neutrino
treatment (Rosswog and Liebend\"orfer 2003).\\ 
The Newtonian self-gravity of the neutron star fluid is calculated using a
binary tree (e.g. Benz et al. 1990). This makes the star less compact and
might -at least quantitatively- influence the accretion process. 
We use a \Pacz-Wiita potential to approximately take into account the
presence of general relativistic effects around the black hole such as the
presence of a last stable orbit. Clearly, the use of the \Pacz-Wiita
potential is not a complete substitute for fully fledged general relativistic
hydrodynamic simulations around a Schwarzschild black hole. It has, however,
turned out to be astonishingly accurate: test-particle orbits with $r<$ 6
M$_{\rm BH}$ (geometrical units with G=c=1 are used throughout the paper) are
unstable and orbits with $r<$ 4 M$_{\rm BH}$ are unbound, i.e. the radius of
marginally stable orbit is located at $R_{\rm isco}$= 6 M$_{\rm BH}$, and the
radius of the marginally bound orbit is R$_{\rm mb}$= 4 M$_{\rm BH}$. Direct
comparisons with general relativistic solutions in a Schwarzschild space time
show that the pseudo-potential is able to capture the essentials of general
relativity and can reproduce accretion disk structures to an accuracy of
better than 10 $\%$ (see e.g. Artemova et al. 1996). The pseudo-potential
does, however, not 
prevent velocities from becoming larger than unity. Abramowicz et al. (1996)
have suggested a rescaling of the velocities found using Newtonian models
together with the \Pacz-Wiita potential (i.e. in our case the numerical
velocities): $v_{\rm num}= v_{\rm phys}\cdot \tilde{\gamma}$, where
$\tilde{\gamma}= (1-(v_{\rm phys})^2)^{-1/2}$. They found this rescaling to
reproduce the exact relativistic velocities to better than 5 \%. This could be
used, for example, to calculate Doppler shifts, this rescaling is however
nowhere used in this paper.\\  
The build up of the binary tree from scratch is computationally expensive,
therefore we do not remove the particles that have crossed our inner
boundary (located at R$_{\rm bd}$= 3 M$_{\rm BH}$) at each time step. They are
only removed when a dump is written, i.e. every $2.5 \cdot 10^{-5}$ s which is
as short as 1/12 of the neutron star dynamical time scale and therefore does not
lead to any artifacts (this has been confirmed by removing particles every
time step). To avoid numerical problems with the singularity of the
\Pacz-Wiita potential we have extended it smoothly with a polynomial (see
Appendix A and Fig. \ref{denom}). Note that all particles that have ever encountered a deviation
from the \Pacz-Wiita potential inside R$_{\rm bd}$ will be removed at the next
dump.\\
We have implemented two different time integration schemes with individual
time steps: a second-order Runge-Kutta-Fehlberg method (Fehlberg 1968) and a
second-order predictor-corrector scheme. Both of these schemes have been
tested extensively against a Runge-Kutta-Fehlberg scheme with a global time
step. As the results were nearly indistinguishable the simulations were run
with the predictor-corrector method as it only requires one force evaluation
per time step. The most time consuming part of these simulations is the
inspiral where (due to the stiff neutron star EOS and the resulting flat
stellar density profile) practically all particles have to be evolved on the
shortest time step. The gain in speed for the presented calculations is
moderate (factors of a few), but in other test problems the code was faster by
more than two orders of magnitude.\\
We construct our initial neutron stars by solving the Lane-Emden-equations
for our EOS with the additional constraint of (cold) $\beta$-equilibrium. This
yields stellar radii of about 16 km for a 1.4 \Msun neutron star.
All SPH-particles have the same mass to avoid numerical noise arising from
interactions of particles with unequal masses. Prior to setting up the binary
system the neutron stars are relaxed carefully by applying an additional,
velocity proportional damping forces in the equations of motion so that the
SPH-particles can settle into their equilibrium positions.
Once such an equilibrium is obtained the NS-BH-binary system is set up, either
with a tidally locked or a non-spinning neutron star. The locked systems 
are constructed  accurately using the hydrocode itself (see Rosswog et
al. 2004). The non-spinning neutron stars are set on orbit with an angular
frequency determined by the force balance between centrifugal and
gravitational forces. In all cases the radial velocity component of a point
mass binary of the considered separation is added.\\
The simulations presented here use up to 3$\cdot 10^{6}$ SPH particles and
are  currently the best resolved models of neutron star black hole
encounters (Rosswog et al. (2004) used up to 10$^{6}$ particles, Lee (2001)
used slightly more than 80000 particles).\\

\begin{table}
\caption{Summary of the different runs. M$_{\rm BH}$: black hole mass in solar
  units;
  $q$=M$_{\rm NS}$/M$_{\rm BH}$; ns spin: C= tidally locked, I= non-spinning
  neutron star; R$_{\rm tid}$: tidal radius;
$a_0:$ initial separation; R$_{\rm isco}$: last stable orbit Schwarzschild
black hole; R$_{\rm MT}$: distance where numerically resolvable mass transfer sets
in; all radii are given in units of km; $\#$ part.: SPH particle number;
  $T_{\rm sim}:$ simulated duration in ms.}

\begin{flushleft}
\begin{tabular}{rccccccrccccccccc} 
run & M$_{\rm BH}$/$q$ & ns spin & $R_{\rm tid}$ & $a_{0}$  & R$_{\rm isco}$ &
R$_{\rm MT}$ & \# part. & $T_{\rm sim}$\\ \hline \\

%
%
I    & 14/0.1   & C &  36.1         &  127.5  &  124.1         &  117 &
570587  &    34.6    \\
%
%
%
%
II  &  14/0.1  & C &   36.1        &   127.5 &  124.1         & 125 & 2971627
&  40.8    \\ 
%
%
%
%
III    & 16/0.0875& C &   37.7        &  145.5  &  141.8         & 122 &
570587  &    78.1  \\
%
%
%
%
IV    & 16/0.0875& C &   37.7        &  145.5  &  141.8         & 123 &
1005401 &    60.9   \\
%
%
%
V    & 18/0.0778      &  C &  39.3       &  162.0  &  159.6      & 123 &
570587  &   50.4    \\
%
%
%
%
VI    & 20/0.07      &  C &  40.7      &  187.5  &  177.3        & 128  &
1503419  &   179.6 \\ 

\\
%
%
VII    & 14/0.1      &  I & 36.1       &  126.2  &  124.1        & 116  &
1497453  &   49.8 \\
%
%
%
VIII    & 16/0.0875      &  I &  37.7      &  143.9  &  141.8        & 121  &
1497453 &   39.8 \\
%
%
%
IX   & 18/0.0778     &  I &  39.3      &  160.4  &  159.6        & 126  &
1497453  &   205.4  \\
\end{tabular}
\end{flushleft}
\label{runs}
\end{table} 

\section{Results}

\begin{table}
\caption{Results the different runs. M$_{\rm BH}$: black hole mass in solar
  units;
  $q$=M$_{\rm NS}$/M$_{\rm BH}$; 
a$_{\rm BH}= J_{\rm BH}/M_{\rm BH}^2$ is the dimensionless black hole spin
  parameter at the end of the simulation ($J_{\rm BH}$: angular momentum
  transferred into the hole); L$_{\rm \nu, tot}^{\rm peak}$ is the peak
  luminosity (ers/s) of the sum of all neutrino flavors; M$_{\rm ej}$ refers
  to the material (in solar masses) that is dynamically ejected during the 
  merger; E$_{\rm kin, ej}$: kinetic energy in ejecta (erg) at end of
  simulation.} 

\begin{flushleft}
\begin{tabular}{rccccccrccccccccc} 
run & M$_{\rm BH}$/$q$ & a$_{\rm BH}$  & L$_{\rm\nu, tot}^{\rm
  peak}$ & M$_{\rm ej}$ & E$_{\rm kin, ej}$\\ \hline \\

%
%
I    & 14/0.1   & 0.196  & $8\cdot10^{50}$ & 0.20 & $2.7\cdot10^{52}$\\
%
%
%
%
II  &  14/0.1  & 0.200   & $8\cdot10^{50}$  & 0.20 & $2.8\cdot10^{52}$\\ 
%
%
%
%
III    & 16/0.0875& 0.197  & $3\cdot10^{50}$ & 0.15 & $2.1\cdot10^{52}$\\\
%
%
%
%
IV    & 16/0.0875 & 0.197 & $2\cdot10^{50}$ & 0.15  & $2.1\cdot10^{52}$\\
%
%
%
V    & 18/0.0778  & 0.201   & $ < 10^{46}$ & 0.08 & $1.3\cdot10^{52}$ \\
%
%
%
%
VI    & 20/0.07   & 0.198  & $ < 10^{46}$  & 0.01  & $2.2\cdot10^{51}$ \\ 

\\
%
%
VII    & 14/0.1  & 0.203   & $1 \cdot 10^{51}$ & 0.17 & $2.2\cdot10^{52}$  \\
%
%
%
VIII    & 16/0.0875 & 0.206 & $6 \cdot 10^{49}$ & 0.11 & $1.3 \cdot10^{52}$\\
%
%
%
IX   & 18/0.0778 & 0.207  & $ < 10^{46}$  & 0.04 & $5.6 \cdot10^{51}$ \\
\end{tabular}
\end{flushleft}
\label{results}
\end{table}

\subsection{Merger dynamics}
\begin{figure}
    \centerline{\psfig{file=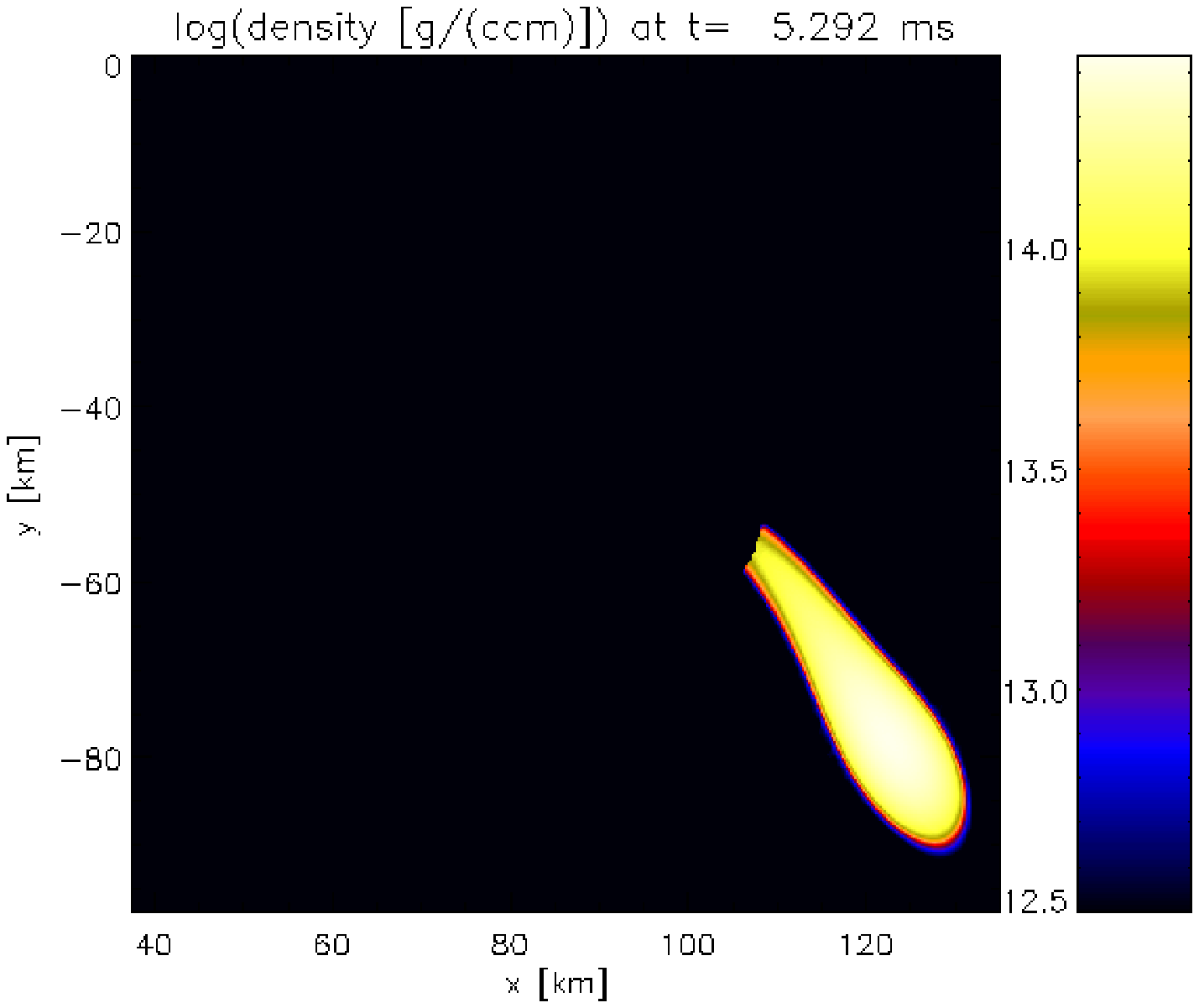,width=7.cm,angle=0}
                \psfig{file=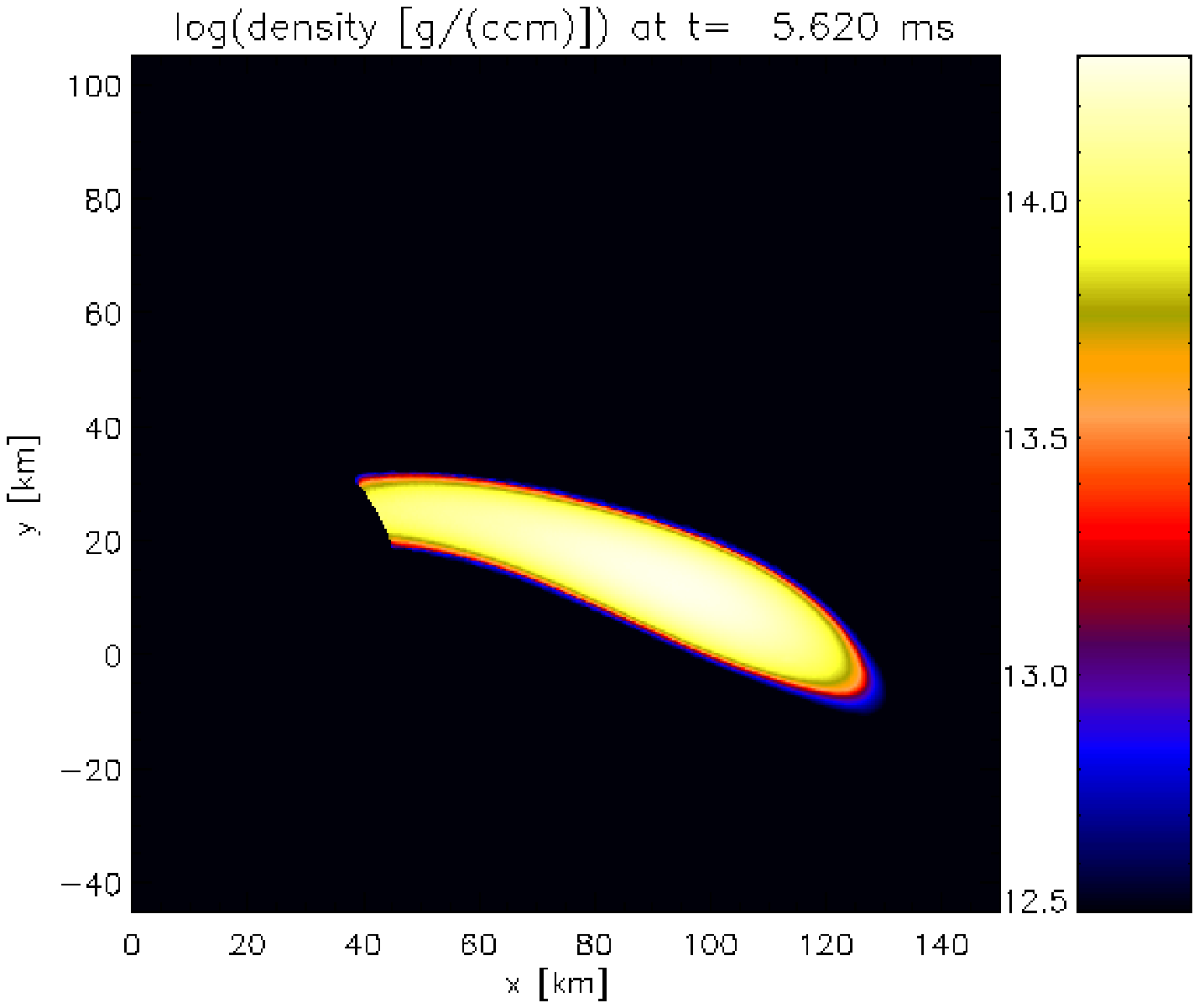,width=7.cm,angle=0}}
    \centerline{\psfig{file=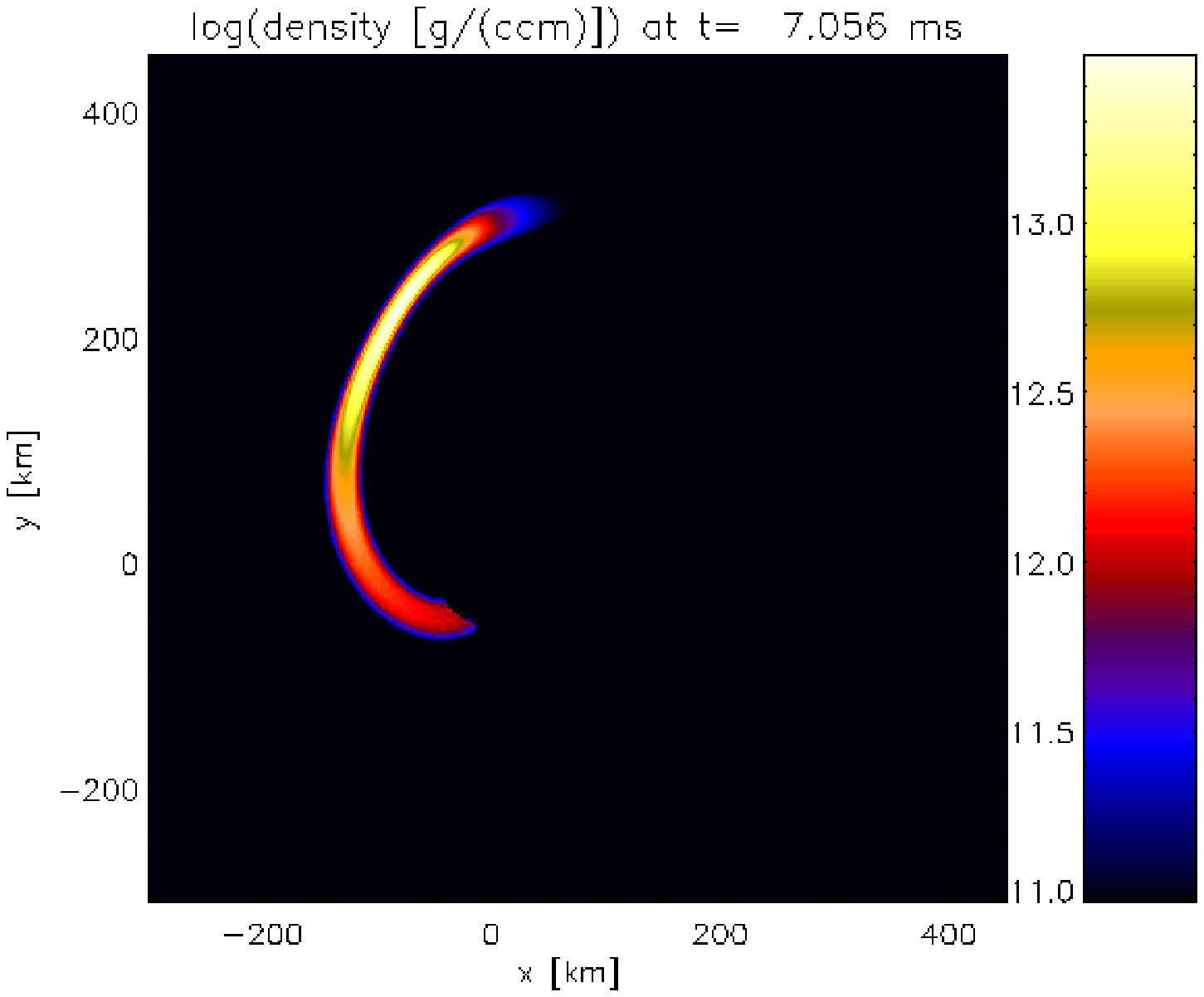,width=7.cm,angle=0}
                \psfig{file=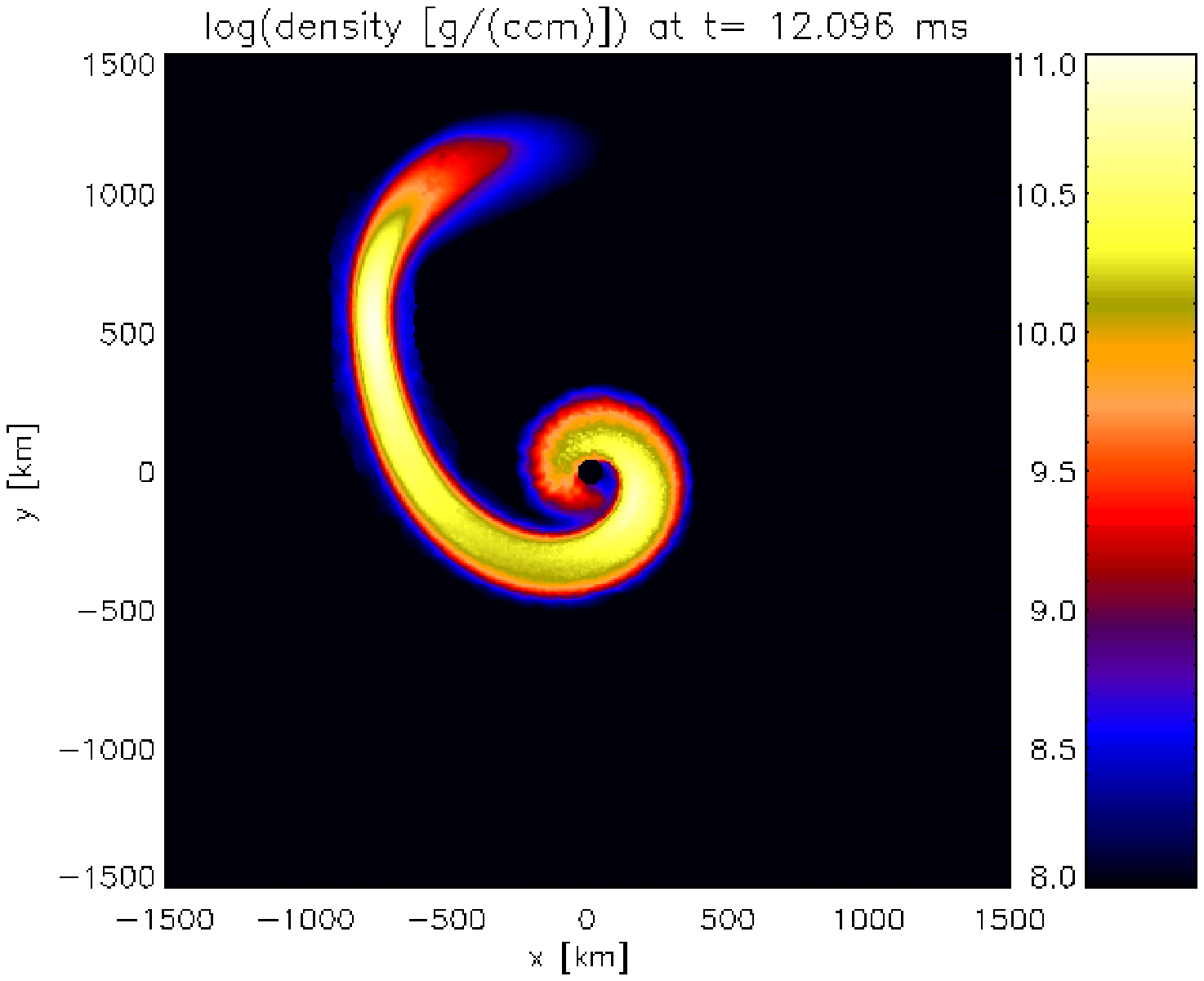,width=7.cm,angle=0}}
    \centerline{\psfig{file=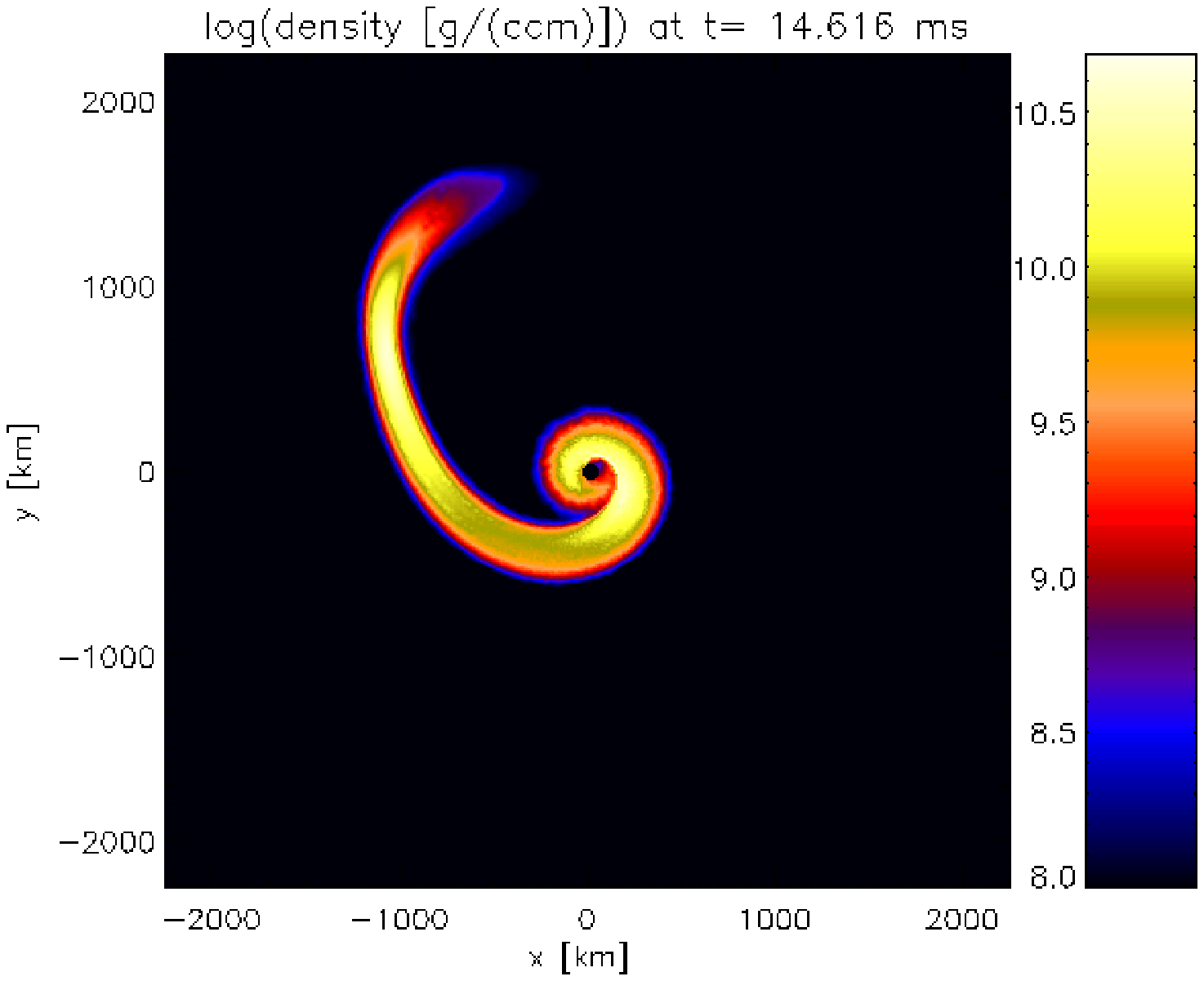,width=7.cm,angle=0}
                \psfig{file=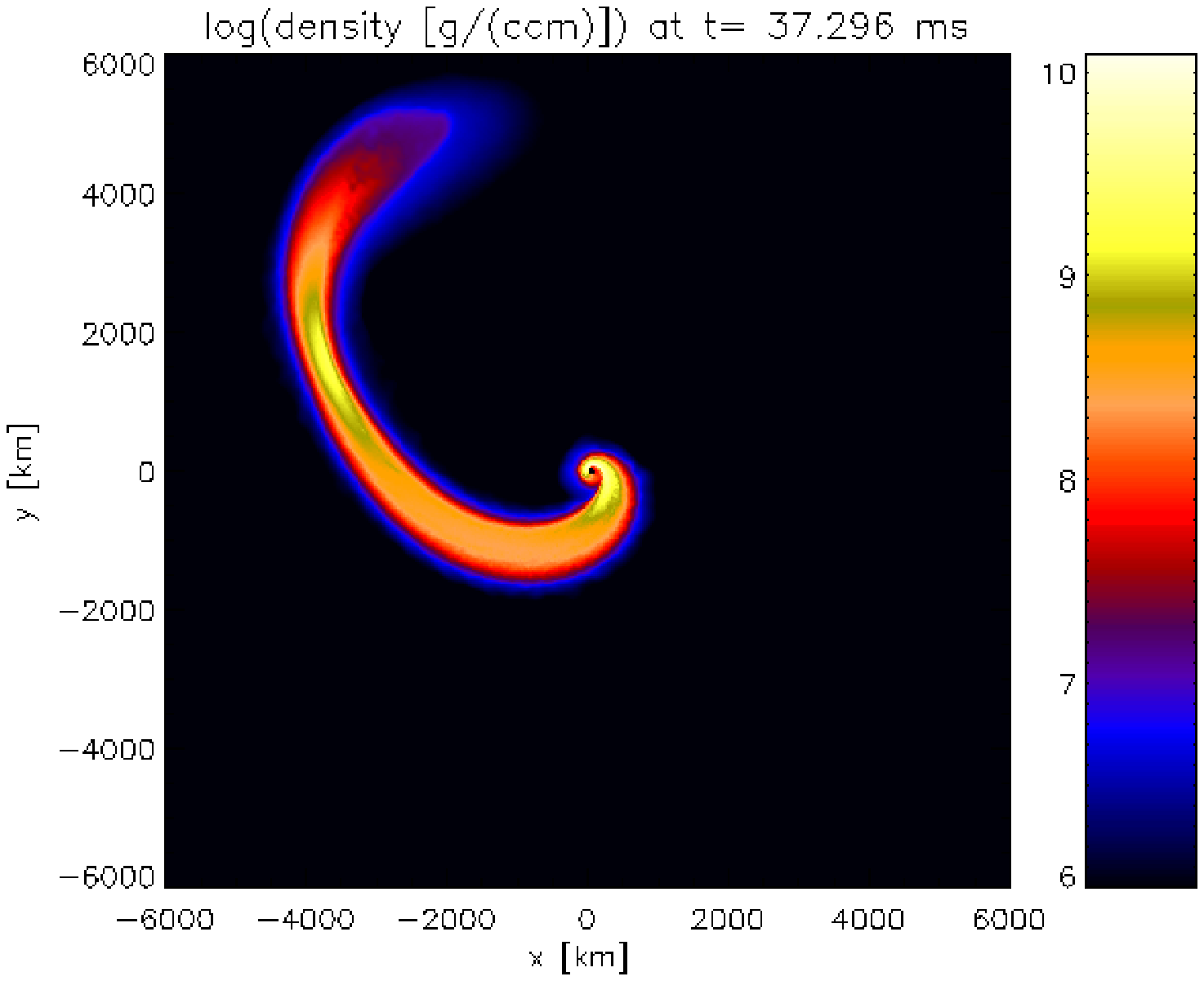,width=7.cm,angle=0}}
    \caption{\label{run II}. Density in the orbital plane of Run II (q=0.1,
                tidally locked neutron star). }
\end{figure}

The dynamics of the coalescence is illustrated in Figures \ref{run II} and
\ref{run VI}, movies can be found at
http://www.faculty.iu-bremen.de/srosswog/movies.html. 
In all of the cases the neutron star is completely disrupted after a large
portion of its mass has been transferred directly into the hole.  The
corresponding peak 
accretion rates exceed 1000 \msun/s for about 1 ms, after this short episode
they drop by at least two orders of magnitude, see panel one in Figure
\ref{mdot}.\\ 
%
%
It is instructive to compare the 14 \Msun case to the corresponding case of
our previous study (Rosswog et al. 2004), where we had used a Newtonian BH
potential. In the purely Newtonian case we found episodic mass transfer with a
low-mass, ``mini neutron star'' surviving throughout the whole simulation or
about eight close encounters. In the \Pacz-Wiita case about 1.15 \Msun
(see panel two in Figure \ref{mdot}) are transferred directly into the hole,
the rest forms a rapidly expanding tidal tail. The tidal tail still contains an
outward-moving density maximum (corresponding to the mini neutron star of the
Newtonian case), but its self-gravity is not strong enough to 
form a spherical object.  The results are well converged, runs I and II show
excellent agreement in the BH masses and peak mass transfer rates. Some minor
deviations are visible at low mass transfer rates (see panel one in
Fig. \ref{mdot} and the distance, $R_{\rm MT}$, where numerically resolvable
mass transfer sets in; see column seven in Table \ref{runs}).\\
The case with 16 \Msun BHs behaves qualitatively very similar to the 14 \Msun
BHs:  slightly more mass (about 1.2 \Msun) is transferred into the hole, the disk is slightly
less massive, hot and dense than the 14 \Msun case. Again, the two different
resolutions yield nearly identical results.\\ 
The systems containing BHs of 18 \Msun or more (runs V, VI and IX) do not form
accretion disks at all. Almost the complete neutron star flows via the inner
Lagrange point directly into the hole, only a small fraction of the star is
spun up enough by tidal torques to be dynamically ejected, see last column in
Table \ref{results}. In these cases the remnant consists of the black hole
(without any accretion disk) and a rapidly expanding, concentric (half-)ring of
neutron-rich debris material (0.08 \Msun for the 18  and 0.01 \Msun for 20
\Msun BH), see Figure \ref{run VI}.\\
The result that it seems to be intrinsically difficult to form promising
accretion disks (at least for the investigated mass ratios) is consistent with
recent estimates of Miller (2005).

\subsection{Disk structure}
A hot and thick accretion disk around a black hole is believed to be an
essential ingredient for a GRB. As described in the previous section, and
as expected from simple analytical estimates, more promising disks form for
larger mass ratios, i.e. smaller BH masses. In the described simulations only
systems with BH masses below 18 \Msun form an accretion disk at all. In the
following the most promising case, run II (M$_{\rm BH}= 14$ \Msun, tidal
locking), is discussed, the $q=0.1$ case with a non-rotating neutron star (run VII)
looks very similar.\\ 
In those cases where a disk forms, substantial parts of it are inside the
innermost stable circular orbit at $R_{\rm isco}= 6$ M$_{\rm BH}$, and are
therefore ``plunging'' with large radial velocities towards the hole which
leads to a substantial consumption of the disks during the simulated time. For
illustration, Fig. \ref{particles_in_disk} shows the distribution of the
SPH-particles at t= 18.396 ms of run II. Shown are the projections of the
particles onto the orbital (=XY-) plane (black) and overlaid are the
projections to the XZ-plane of those particles $j$ with $|y_j|< 150$ km (red)
to show the vertical disk structure. Moreover the locus of the Schwarzschild 
radius and $R_{\rm isco}$ are shown. While the matter cross section far
away form the hole is close to circular, the disk close to the hole is
geometrically thin ($|z|\ll \bar{\omega}= \sqrt{x^2+y^2}$) apart from the
region close to the spiral shock where the accretion stream interacts with
itself. In this regions the disk is slightly puffed up. Due to the plunge
motion only moderate densities ($\log(\rho) < 10.5$ \gcc) and temperatures are
reached (see Fig. \ref{rho_T_runII}). The disk is essentially cold apart form
the spiral shock where the temperatures reach $\sim 2.5$ MeV.\\
The geometrically {\rm thin}, relatively cool disks that we find for black
hole masses between 14 and 16 \Msun is in stark contrast to
the {\rm thick} disks that form in the neutron star merger case (see Figs. 15
and 16 in Rosswog and Davies (2002)).

\subsection{Neutrino emission}

The annihilation of neutrino anti-neutrino pairs is one possibility to deposit
energy in the baryon free region above the hole. In the cases with black hole
masses $\ge 18$ \Msun the neutrino emission negligible ($< 10^{46}$
erg/s). The most promising cases are, as expected, the ones with the lowest
black hole masses. If disks form at all they are completely transparent to
the emerging neutrinos. In lowest mass cases with M$_{\rm BH}= 14$ \Msun (runs
I, II, and VII) we find peak luminosities of $L_{\nu, {\rm tot}}=
L_{\nu_e}+L_{\bar{\nu}_e}+L_{\nu_x} \approx 10^{51}$ erg/s (where the index
$x$ refers to the heavy lepton neutrinos), i.e. our most
promising cases here yield luminosities that are more than two orders of
magnitude smaller than in our neutron star merger calculations where the same
microphysics was used. The  average energies of the emitted neutrinos are
between 12 and 15 MeV. Opposite to the neutron star merger
case the neutrino luminosity does not settle into a stationary state, since
the disk is being consumed on a time scale of tens of milliseconds. In the
tidally locked case we find a single neutrino pulse of about 20 ms duration,
in the case without neutron star spin we find two neutrino pulses separated by
about 15 ms, see Figure \ref{Lnu_tot}.   

\section{Summary and Discussion}
We have performed three-dimensional hydrodynamic simulations of neutron black
hole encounters with mass ratios $q= \frac{M_{\rm ns}}{M_{\rm BH}} \le
0.1$. We used a state-of-the-art temperature-dependent, nuclear EOS, a
detailed, multi-species neutrino treatment and the \Pacz-Wiita
pseudo-potential.\\
We consider all the approximations made to be valid to a high degree. If an
accretion disk forms at all (i.e. for the BH masses $<18$ \Msun) it is of only
moderate density ($\sim 10^{10}$ \gcc) and  
completely transparent to neutrinos. Therefore the neutrino emission
results cannot be influenced by the flux-limited diffusion
treatment. Moreover, the results are numerically converged, different
numerical resolutions yield almost identical
results. The BHs are massive enough to dominate the space-time completely and as
they are spun up to spin parameters of only 0.2 (see Table \ref{results}), we consider the use of
PW-potentials a very good approximation (note that for a=0.2 the event horizon
moves from 2 to 1.98 and the last stable orbit from 6 to 5.33 gravitational
radii; see Fig. \ref{event_horizon}).
\subsection{Implications for nucleosynthesis}
The ejected mass fraction found in these simulations is generally {\em very} large
(see Table \ref{results}), but comparable to the analytical estimates of
Lattimer and Schramm (1974 and 1976), who estimated $0.05\pm0.05$ \msun. 
We had performed previously nucleosynthetic calculations for the decompressed
ejecta in the neutron star merger case (Freiburghaus et al. 1999). As these calculations did not account for
$Y_e$-changes due to weak interactions, $Y_e$ was treated as a free parameter
and was varied in a range that is reasonable for neutron star material (0.05 to 0.20).
For values between 0.08 and 0.15 we found an almost perfect agreement with the
observed, solar system r-process abundances from around Barium up to beyond the platinum
peak. The value of $Y_e$ has basically two effects: i) it determines the
neutron to seed ratio and thus the maximum nucleon number $A$ of the resulting
abundance distribution and ii) the location of the reactions in the N-Z-plane,
the so-called r-process path, which is, in this case, very close to the
neutron drip-line. This path determines the nuclei that are involved and
therefore the nuclear energy release and the corresponding $\beta$-decay time scales. 
The initial neutron star material will be decompressed via the expansion and will
therefore continuously change its composition. Starting from very large 
initial nuclei with hundreds of nucleons (the details of which are determined
by the distributions of density, temperature and electron fraction) these
transmutations will release nuclear binding energy until a final, $\beta$-stable
composition has been reached. Figure \ref{Ye_dist} shows the snapshot distribution of
$Y_e$ within the debris of run II at t= 18.396 ms, in Figure \ref{Ye_binned}
$Y_e$ has been binned with the ejecta mass for the NSBH cases with q=0.1. Generally, 
as the only parts that have ever reached high temperatures disappear very
quickly into the hole (see Fig. \ref{rho_T_runII}), the $Y_e$ of the ejected
material reflects closely its initial value inside the neutron star.
The found $Y_e$s are generally very close to what was found in the neutron
star merger case to give excellent results (compare Freiburghaus et al. 1999,
Fig. 4; note, however, that due to the somewhat different expansion
velocities, the interesting $Y_e$ values may be slightly different in the NSBH
case). To determine the exact elemental yields from these mergers is beyond
the scope of the current work and is left to future investigations.\\  
For the neutron star merger case Argast et al. (2004) found that a conflict
with the observed element ratios in metal-poor halo stars may arise if neutron
star mergers were the {\em dominant} r-process source. Similar questions may
be raised for the NSBH case. One possibility to avoid the problem could be
that NSBH coalesce much less frequently than DNS. This would be consistent
with the non-observation of any NSBH-system (currently 8 DNS systems are
known, see Stairs 2004) and the result of recent studies (Pfahl et al. 2005)
that estimate the NSBH number in the Galactic disk to be less than 0.1-1\% of
the number of DNS.\\ 
We want to point out, however, that the ejecta found in our simulations exhibit
very large radial velocities of $\approx 0.5 c$. As most neutron star black
hole binaries are expected to merge in the outskirts of galaxies (see e.g. Perna and
Belczynski 2002) the chances are large that these high-velocity ejecta do not
end up within the host galaxies, but rather enrich the intergalactic medium
with very heavy, probably high-mass r-process elements. This interesting issue
definitely requires further consideration in the future. 
\subsection{Gamma-ray Bursts}
The formation of a massive accretion disk around a black hole is thought to be 
a vital ingredient for a GRB central engine. None of the systems investigated
in this study yields disks that are promising as GRB engines. As the tidal
radius only grows $\propto$ M$_{\rm BH}^{1/3}$ while the gravitational radius
of the BH scales $\propto$ M$_{\rm BH}$, low mass black holes will easier form
promising disks. In the simulations presented here, disks only form for BHs
with masses below 18 \msun. 
As mentioned previously, these disks are essentially cold apart from a spiral
shock with temperatures of $\approx 2.5$ MeV. Contrary to the neutron star
merger case, these disks are geometrically {\em thin}, of
only moderate density and completely transparent to neutrinos. The peak densities in the
shock are $\log(\rho) \approx 10.5$ (cgs) and lower elsewhere (see
Fig. \ref{rho_T_runII}). This is to be compared with peak values of
$\log(\rho) \approx 12$ (cgs) and $T\approx 4$ MeV in the neutron star merger
case (Rosswog and Davies 2002). Moreover, the disk is drained 
substantially on the simulation time scale of a few 10 ms.\\
In the most promising cases the peak luminosities are about $10^{51}$ erg/s
(see Table \ref{results}, Fig. \ref{Lnu_tot}). This is 
more than two orders of magnitude below the typical neutron star merger case (Rosswog and
Liebend\"orfer, 2003). The hot debris of a double neutron star coalescence
deposits $\sim 10^{48}$ erg/s via $\nu\bar{\nu}$-annihilation in baryon-devoid
regions and drives in this way a relativistic, bipolar outflow (Rosswog and
Ramirez-Ruiz 2002, Aloy et al. 2005). The surrounding baryonic material collimates
these jets into a small fraction of the solid angle so that they appear as
isoptropized $\sim 10^{50}$ erg (Rosswog and Ramirez-Ruiz 2003). 
As the neutrino annihilation rate scales roughly with the neutrino luminosity
squared, the most promising cases of this study (q= 0.1) will provide $L_{\rm
  jet, NSBH} \approx 3\cdot10^{48} \cdot \left( \frac{10^{51}}{2\cdot 10^{53}}\right)^2 {\rm 
  erg/s} \approx 10^{43}$ erg/s for about 10 ms. This estimate is rather on
the optimistic side as, due to the thin disk geometry in the NSBH case, the
probability for neutrinos to collide close to head-on (where the
annihilation cross-section is maximal) is severely reduced. Moreover, such
disks will not provide powerful winds from neutrino ablation to collimate a
possible outflow. Therefore, the signal resulting from
neutrino annihilation may be even less luminous. General relativistic effects
like the bending of neutrino trajectories and the redshift of the neutrino energy
may influence the result, but their influence goes in opposite directions
and will change the non-relativistic results by only a factor of about two
(Asano and Fukuyama 2001).
Recently, Ramirez-Ruiz and Socrates (2005) have argued that neutrino spectra
substantially different from black body may increase the overall explosion
efficiency by more than an order of magnitude. But even if we stretch all of
the available parameters to their most optimistic limits, e.g. assuming one
order of magnitude more efficiency from neutrino spectra and an (ad hoc) extremely small
beaming fraction $\zeta=\Omega/2 \pi \sim 10^{-3}$, it is hard to see how the
apparent isotropized luminosity should exceed, say, $10^{47}$ erg/s (for
about 10 ms) even in the most promising cases of this study.\\ 
Another agent that can possibly feed energy into collimated outflow is the
magnetic field. Being naturally endowed by strong magnetic fields to start
with the shredding of a neutron star could be expected to be a good candidate
for magnetic field amplification. This may occur via various pathways, the
fastest of which is probably the magneto-rotational instability
(Velikhov 1959, Chandrasekhar 1991, Balbus and Hawley 1998 and references
therein). Fig. \ref{Beq} shows the evolution of the disk-averaged magnetic
field, 
\begin{equation}
\langle B^{\rm eq} \rangle = \sqrt{8\pi} \frac{\sum_i m_i \sqrt{\rho_i} c_{s,i}}{\sum_i m_i} 
\end{equation}
making the extreme assumption that the field can reach equipartition
field strength. Here, the index $i$ runs over the particles inside 600 km (see
Figure \ref{particles_in_disk}) around the hole, $m_i$ are the masses,
$\rho_i$ the densities and $c_{s,i}$ the sound velocities of the
SPH-particles. Under this assumption an average field strength slightly in excess of
$10^{14}$ G would result. Making the most optimistic assumptions, we will
estimate the properties of the possibly resulting fireball (see Piran 1999 for
a fireball review). The fireball energy is $E \sim \frac{\langle B^{\rm eq}
\rangle^2}{8 \pi} V_{\rm ISCO} \approx
4\cdot10^{48} {\rm erg} \; B_{14}^2 (\frac{M_{\rm BH}}{15 {\rm
    M}_{\odot}})^3$. This yields an initial fireball temperature of $T_0
\approx 3.3 \;{\rm MeV} \; \left( \frac{E}{4\cdot10^{48} {\rm erg}}\right)^{1/4}
\left(\frac{M_{\rm BH}}{15{\rm M}_{\odot}} \right)^{-3/4}$. The critical value
$\eta_b$, where $\eta= E/M$ with $E$ and $M$ energy and mass of the fireball,
is given by (Piran 1999), $\eta_b \approx 6000 \; (\frac{E}{4\cdot10^{48} {\rm
  erg}})^{1/3} \left(\frac{M_{\rm BH}}{15{\rm M}_{\odot}}
\right)^{-2/3}$. Given the fact that the resulting disk is very thin and
neutrino ablation not important, we consider it plausible that the fireball is
in the electron opacity dominated range, $\eta_b < \eta < \eta_{\rm pair}
\approx 5\cdot10^{8} \; (\frac{E}{4\cdot10^{48} {\rm erg}})^{1/2} (\frac{M_{\rm
  BH}}{15{\rm M}_{\odot}})^{-1/2}$. The fireball will thus become transparent to photons before it
reaches the matter dominated stage. As the Lorentz-factor grows $\gamma \propto R$
and the fireball temperature drops $T \propto 1/R$, the observed temperature,
$T_{\rm obs} \propto \gamma T$, will be that of the initial fireball. This
will yield a black body gamma-ray pulse with a temperature of about 3.3
MeV. Performing similar estimates for the fireball from neutrino annihilation yields
a thermal pulse of about 2 MeV. SWIFT will have difficulties to detect it, but
GLAST should be able to see a pulse of several tens of milliseconds out to a
redshift of $z \sim 0.1$.\\
As mentioned previously, these estimates are based on most optimistic assumptions. 
As the material plunges into the hole within about one orbit once it has
passed R$_{\rm ISCO}$, any instability will have difficulties to amplify
the field by large factors. The expected thermal precursor pulse
may therefore be even weaker than estimated above.\\
In summary, none of the investigated systems seems to be an obvious candidate
for the central engine of the observed short-hard GRBs.
Neutron star black hole mergers may still be able to produce GRBs. Obviously,
smaller mass black holes are more promising. Technically, however, they are
much more challenging as the neutron star has a mass comparable mass to the
black hole and space-time is far from the Schwarzschild solution. Therefore,
neither the use of pseudo-potentials nor simulations on a fixed background
space-time are justified, here fully relativistic 3D simulations of the full
merger process are required.\\
The spin of the black hole may provide a possibility to launch GRBs even from
more massive black holes. If they are spinning very rapidly from the
beginning and are spun up further during the merger to values close to the maximum
spin parameter of $a=1$ then both the position of the last stable orbit and of
the event horizon move to 1 M$_{\rm BH}$, see
Fig. \ref{event_horizon}. Therefore, much higher temperatures and 
densities can be reached in the inner disk regions. The fraction of viable GRB
candidates among the NSBH systems obviously depends on their mass and initial
BH spin distribution. Both quantities are not observationally known and their
predictions via population synthesis models are plagued by large
uncertainties. It is however clear, that not the whole available parameter
space will provide conditions that are favorable to launch GRBs. This
fact and the non-observation of even a single neutron star black hole system
(while eight double neutron star systems are observed) makes it plausible that
the observed short-hard GRBs are dominated by double neutron star
coalescences.\\  
\subsection{Electromagnetic transients}
In the investigated NSBH mergers we find very large amounts of radioactive
material being  
ejected. The radioactive decay of this debris material would, like in a type Ia 
supernova, power the post-coalescence lightcurve. The bulk of ejecta material stems
from the initial nucleon fluid of the neutron star and will end up in large,
possibly r-process, nuclei.  If we assume a typical nuclear binding energy of
$\sim$ 8 MeV per nucleon for the final nuclei, the decompression of this material
will release, depending on the mass ratio of the NSBH system, between  $\sim 1.6
\cdot 10^{50}$ and $\sim 3.2 \cdot 10^{51}$ erg from radioactive decays. The
details of the resulting lightcurve will depend on the nuclear binding energies and decay time scales
and their competition with the expansion of the material. As these nuclei will
be extremely neutron rich (see Fig. \ref{Ye_binned}) and far away from the valley of
$\beta$-stability no experimental data for their half-lives are available and
one would have to resort to theoretical predictions. For a simple
estimate of the resulting electromagnetic transient from these radioactive ejecta we use the simple
analytical estimate of Li and \Pacz (1998) who assumed spherical symmetry and
a uniform distribution of nuclear lifetimes in logarithmic time intervals. Scaling
their results with the numbers found in our simulations we find that the peak
luminosity of $\sim 6\cdot10^{44}$ erg/s should be reached about 3 
days after the coalescence (the 'banana-like' geometry may lead to a peak that occurs
somewhat later than in the spherical model of Li and \Pacz). The effective
temperature at peak will be $\sim 10^4$ K and result in an intensity maximum
in the optical/near infrared band of the spectrum.\\
It is, however, worth pointing out that this result relies on radioactive
nuclei still being present when the ejecta become optically thin, which
depends on the details of the nuclear reaction path. It may also be
possible that the decays have ceased already by that time and the result is an
essentially 'dark explosion'. This interesting topic definitely deserves
further examination.\\
In summary, the merger of a neutron star with a black hole of a mass beyond 14
\Msun may, rather than producing a gamma-ray burst, yield a supernova-like
transient with a thermal precursor pulse peaking in the gamma-ray band. Such a
precursor of $\sim 10$ ms duration should be detectable with GLAST out to a
redshift of $z \sim 0.1$. Follow-up observations should then be able to detect
a few days later the supernova-like transient.\\

\noindent Movies of the simulations can be found under \\
http://www.faculty.iu-bremen.de/srosswog/movies.html.

\acknowledgments
\noindent It is a pleasure to thank Marek Abramowicz, Jim Lattimer, William Lee, Andrew
MacFadyen, Enrico Ramirez-Ruiz,
Roland Speith and Christophe Winisdoerffer for useful discussions and the INFN
in Catania and the IAS in Princeton for their hospitality. The calculations
reported here have been performed on the JUMP supercomputer of the
H\"ochstleistungsrechenzentrum J\"ulich. 

\newpage

{\bf Appendix A: \Pacz-Wiita potential with smooth extension}\\
Here we show the form of the potential that is used in the
simulations. To numerically avoid the singularity at the Schwarzschild-radius
we extend the \Pacz-Wiita acceleration denominator, $D_{\rm PW}=
\frac{1}{(r-R_s)^2}$, smoothly down to vanishing radius with a polynomial,
$D_{\rm pol}$, that posesses the following properties: \\
i) it matches $D_{\rm PW}$ at a transition radius, $R_t$,: $D_{\rm PW}(R_t)=
D_{\rm pol}(R_t)$, \\
ii) the derivatives match smoothly: $\frac{d D_{\rm PW}}{dr}(R_t)= \frac{d D_{\rm
    pol}}{dr}(R_t)$, \\
iii) its derivative vanishes in the origin $\frac{d D_{\rm pol}}{dr}(0)= 0$.\\
Using $A= R_t^{-1}\cdot(R_s-R_t)^{-3}$ and $B= \frac{R_s - 2
  R_t}{(R_s-R_t)^3}$ the force denominator is expressed as

\[
D = \left\{ \begin{array}{r@{ }l}
                         & A r^2 + B  \quad \mbox{for $r < R_t$}\\
                         & \frac{1}{(r-R_s)^2}   \qquad \mbox{for $r \ge  R_t$.}
                                     \end{array}
                             \right  .\]

The form of $D$ is shown in Fig. \ref{denom}, we always use $R_t= 3$ M$_{\rm BH}$. Note that every
particle that has ever been inside $R_t$ is removed at the next dump step (the
time between two subsequent dumps is a small fraction ($\approx 1/12$) of the
neutron star dynamical time).\\

\newpage


%
%

\clearpage

\clearpage
\begin{figure}
    \centerline{\psfig{file=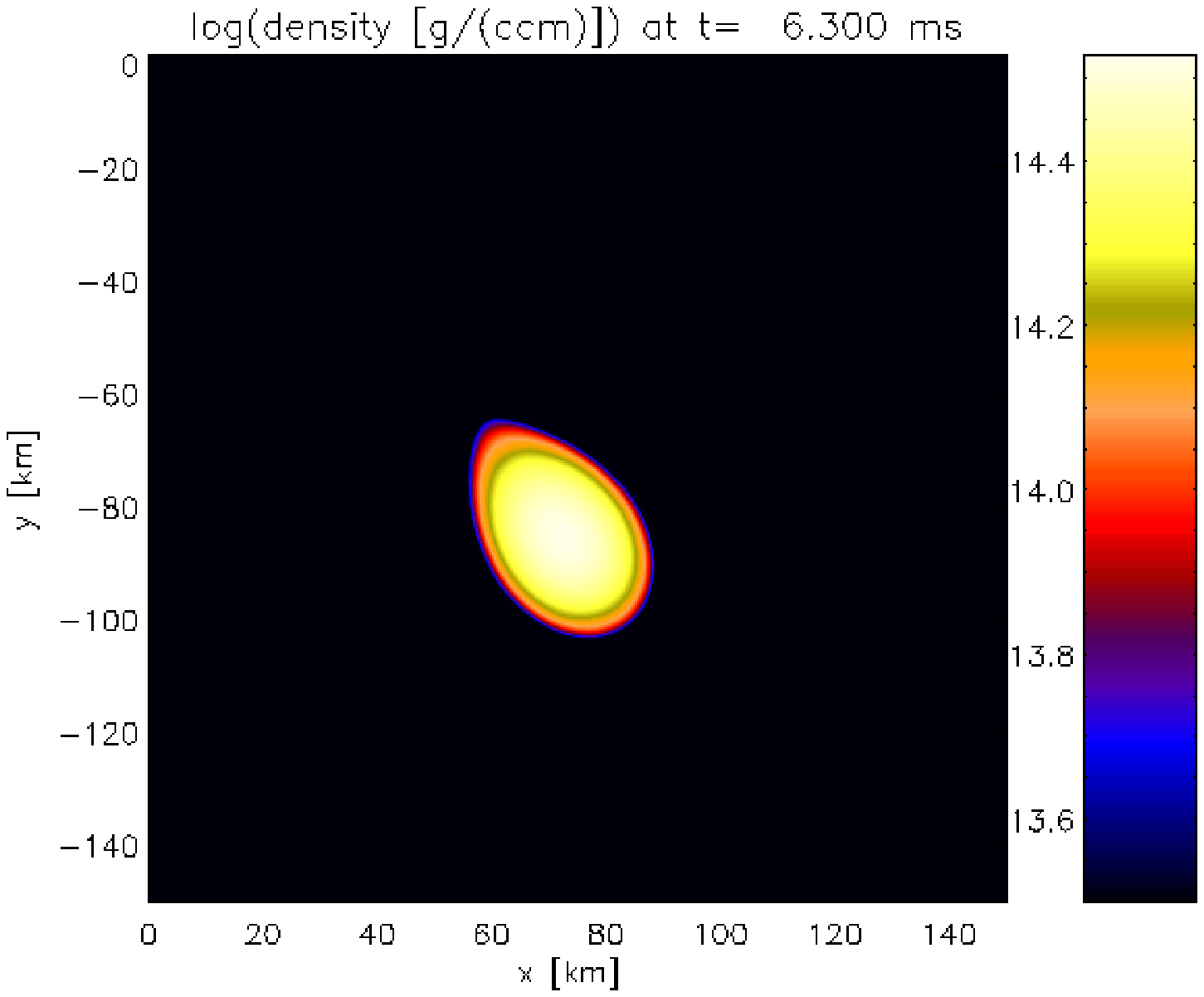,width=7cm,angle=0}
                \psfig{file=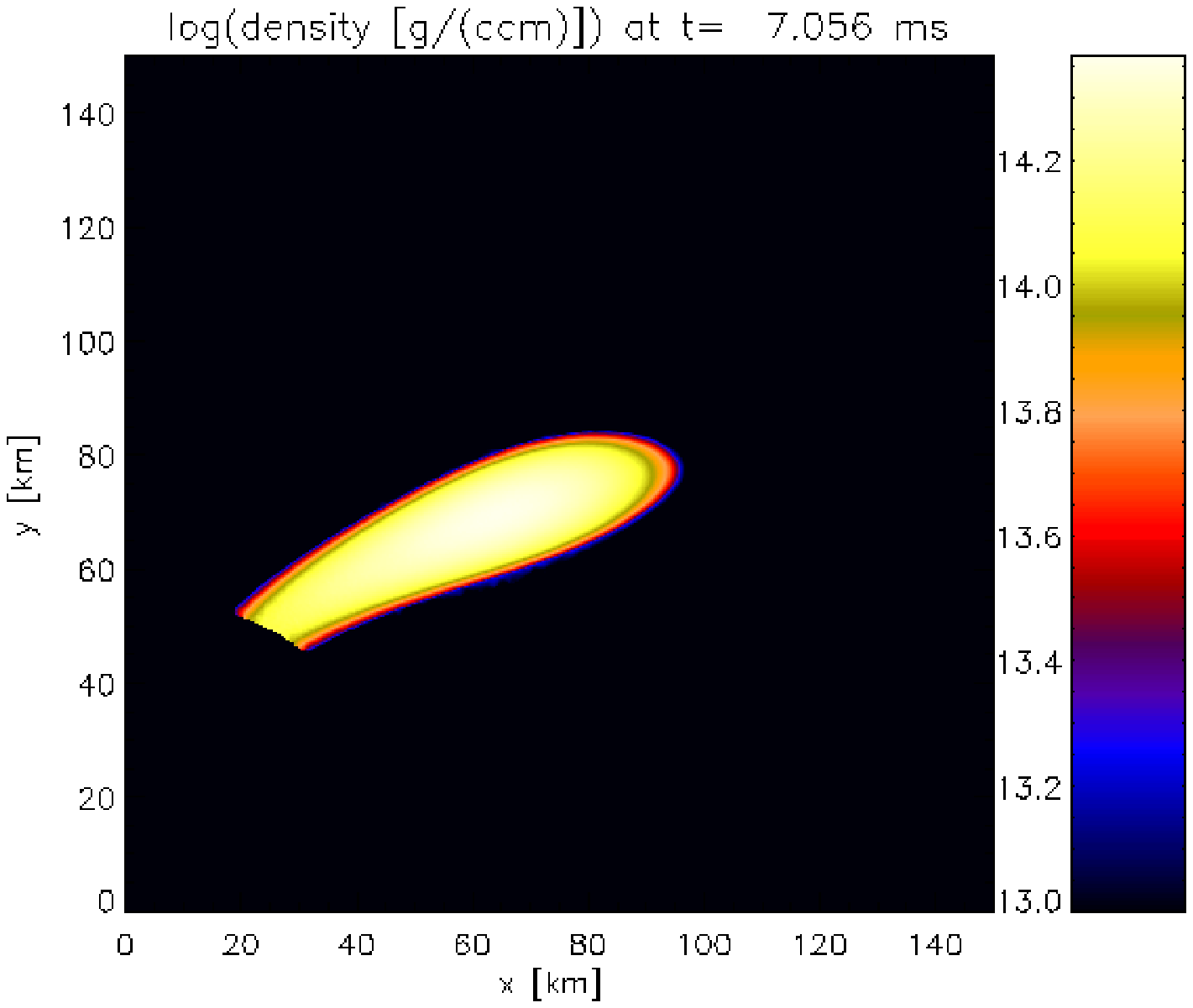,width=7cm,angle=0}}
    \centerline{\psfig{file=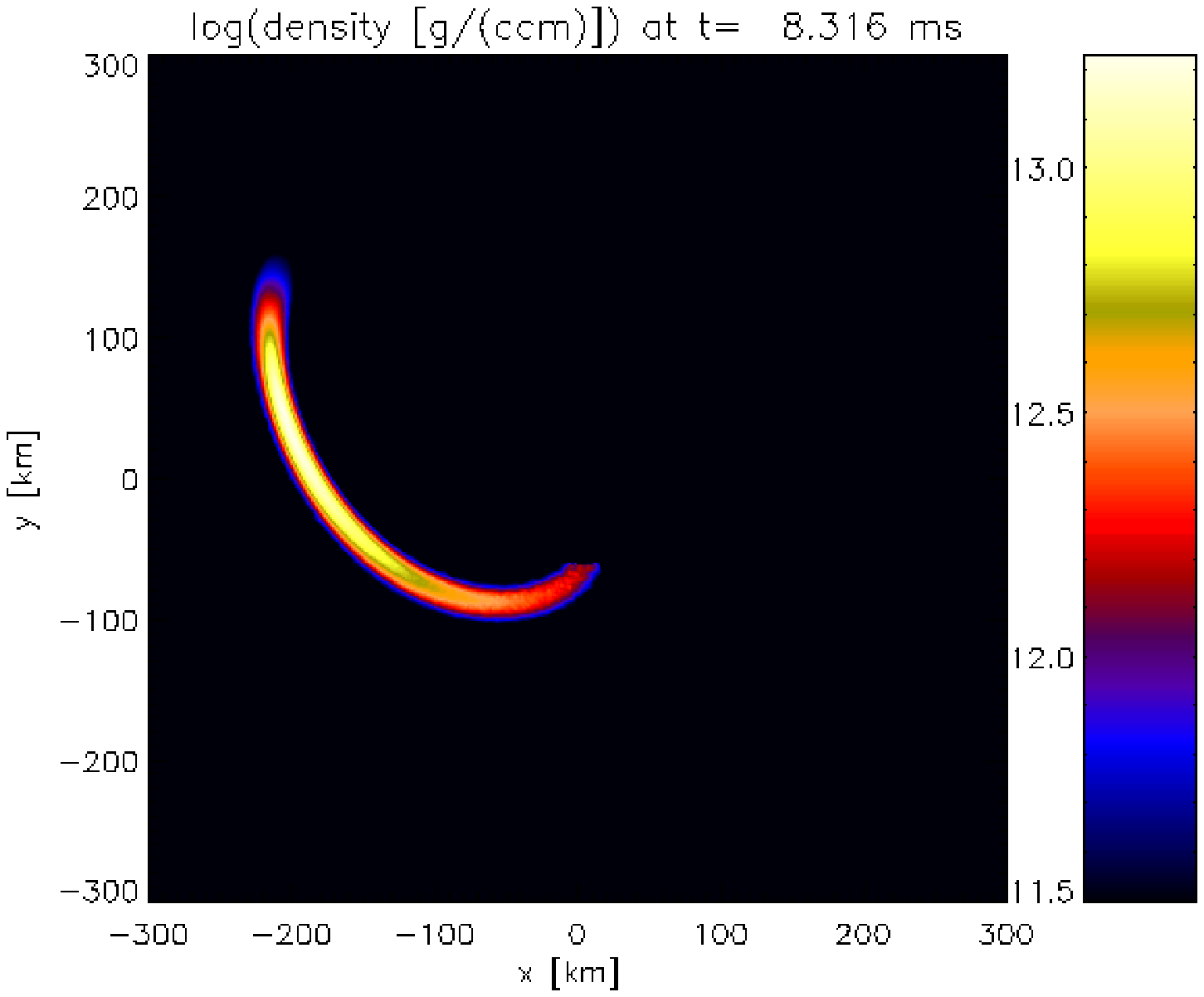,width=7cm,angle=0}
                \psfig{file=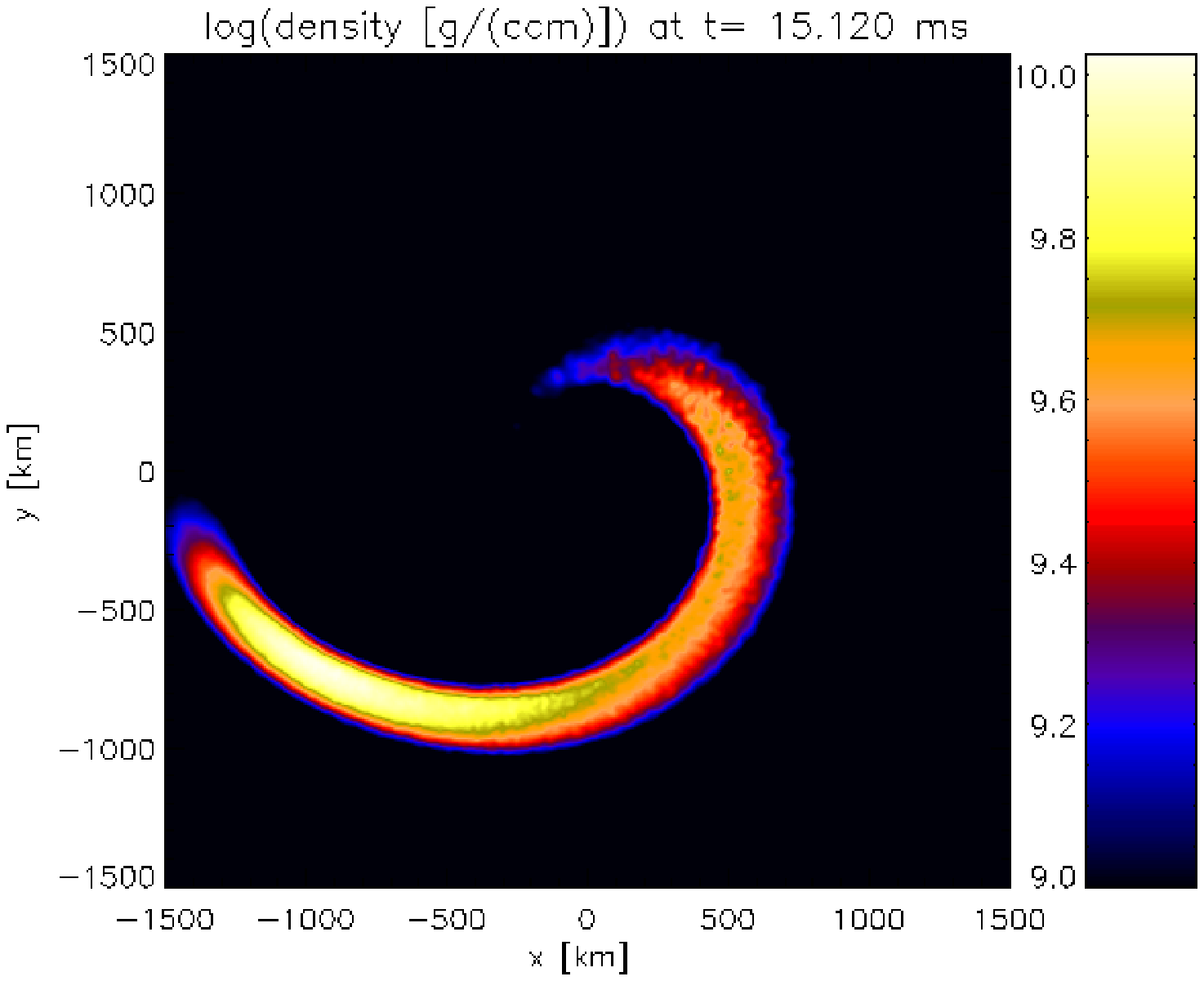,width=7cm,angle=0}}
    \centerline{\psfig{file=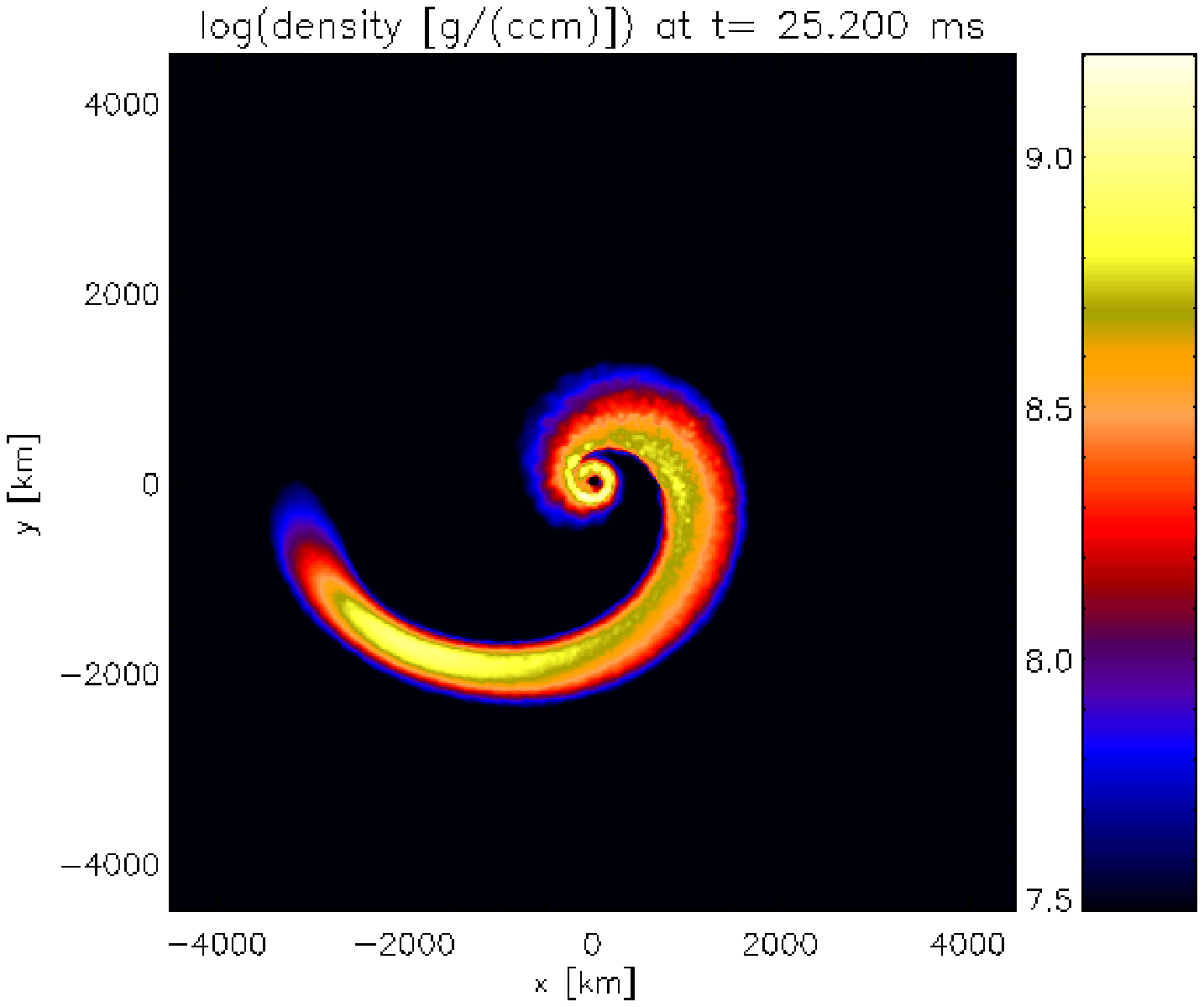,width=7cm,angle=0}
                \psfig{file=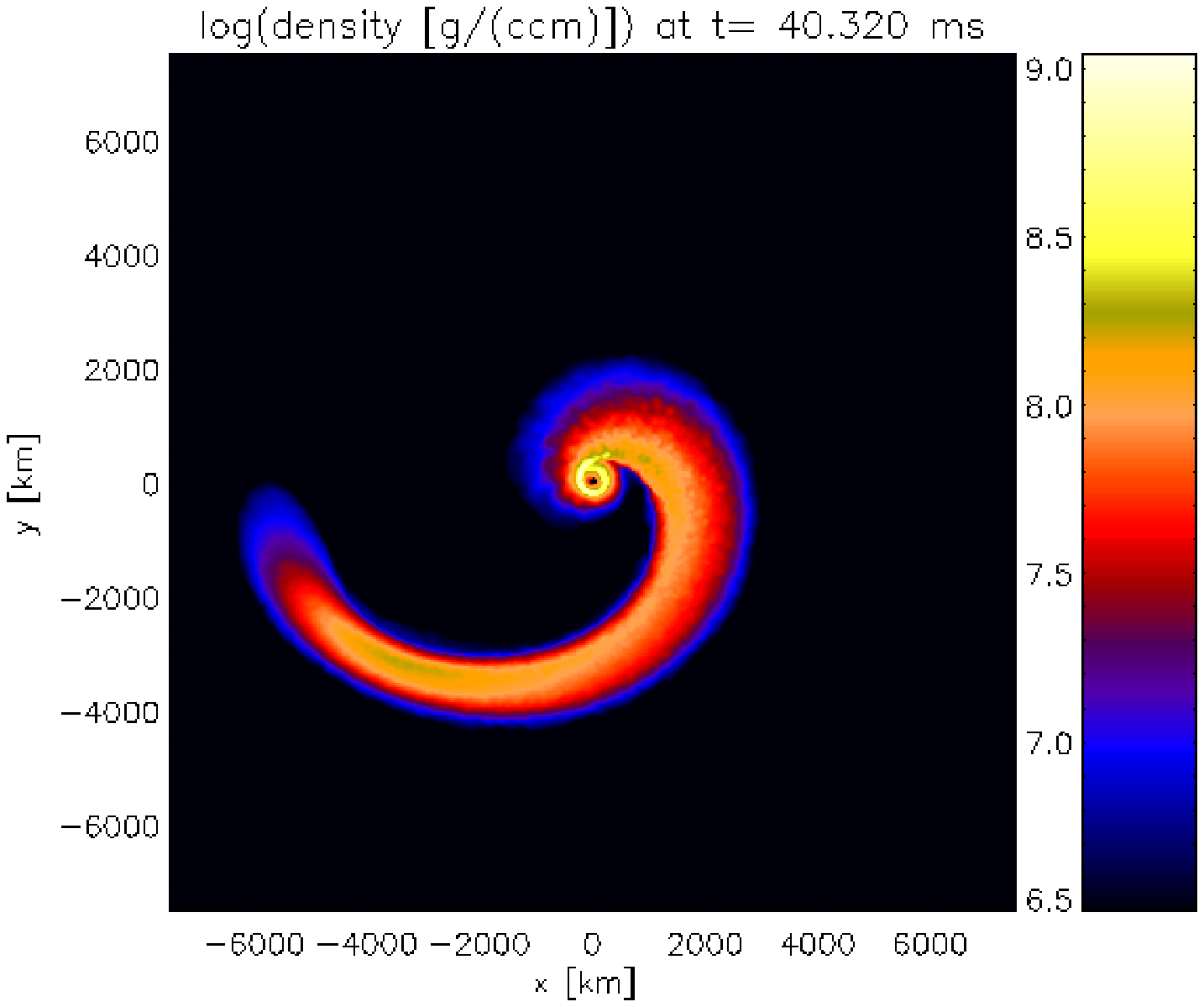,width=7cm,angle=0}}
    \caption{\label{run IV}. Density in the orbital plane of Run IV (q=0.0875,
                tidally locked neutron star). }
\end{figure}

\clearpage
\begin{figure}
    \centerline{\psfig{file=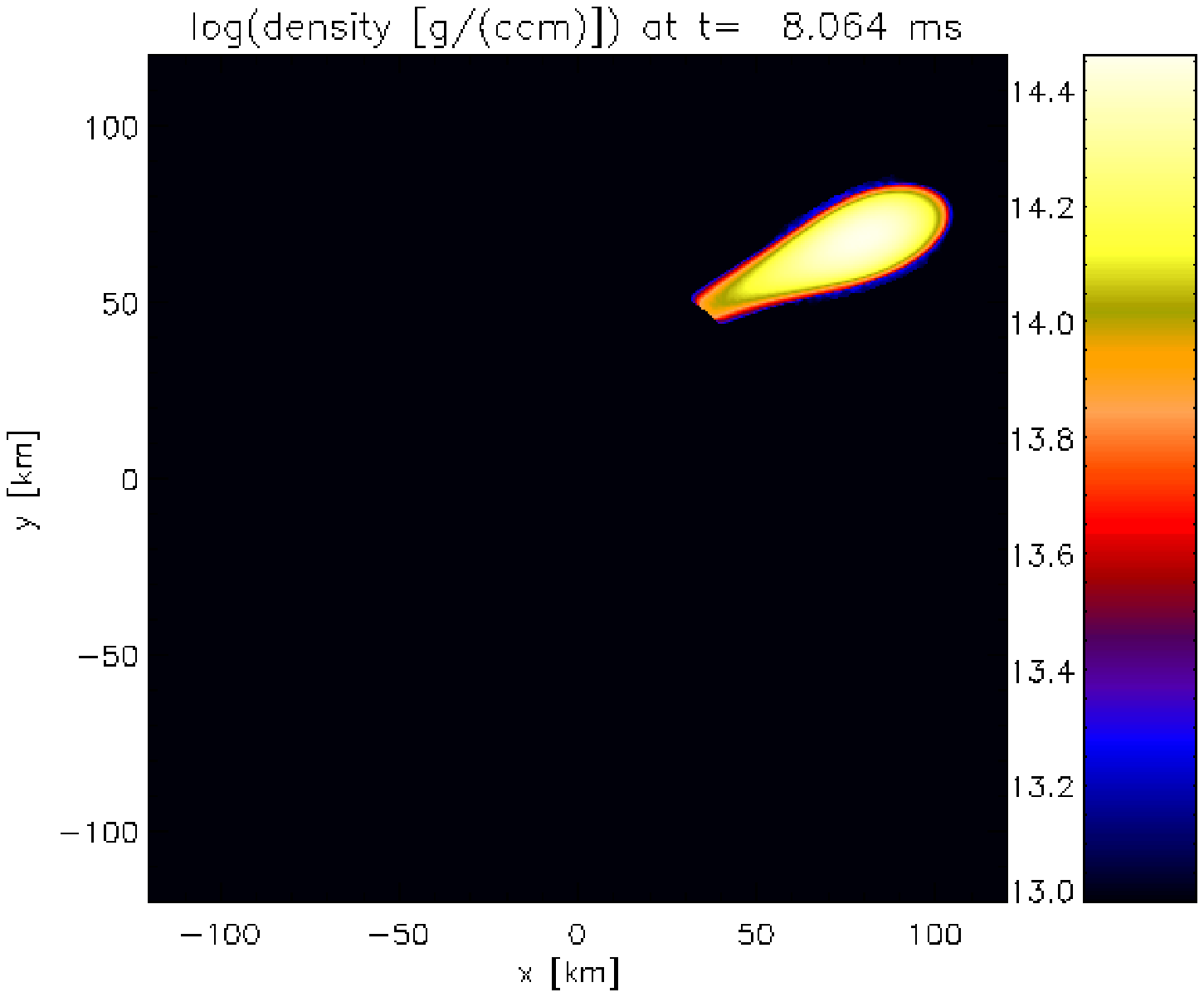,width=7.cm,angle=0}
                \psfig{file=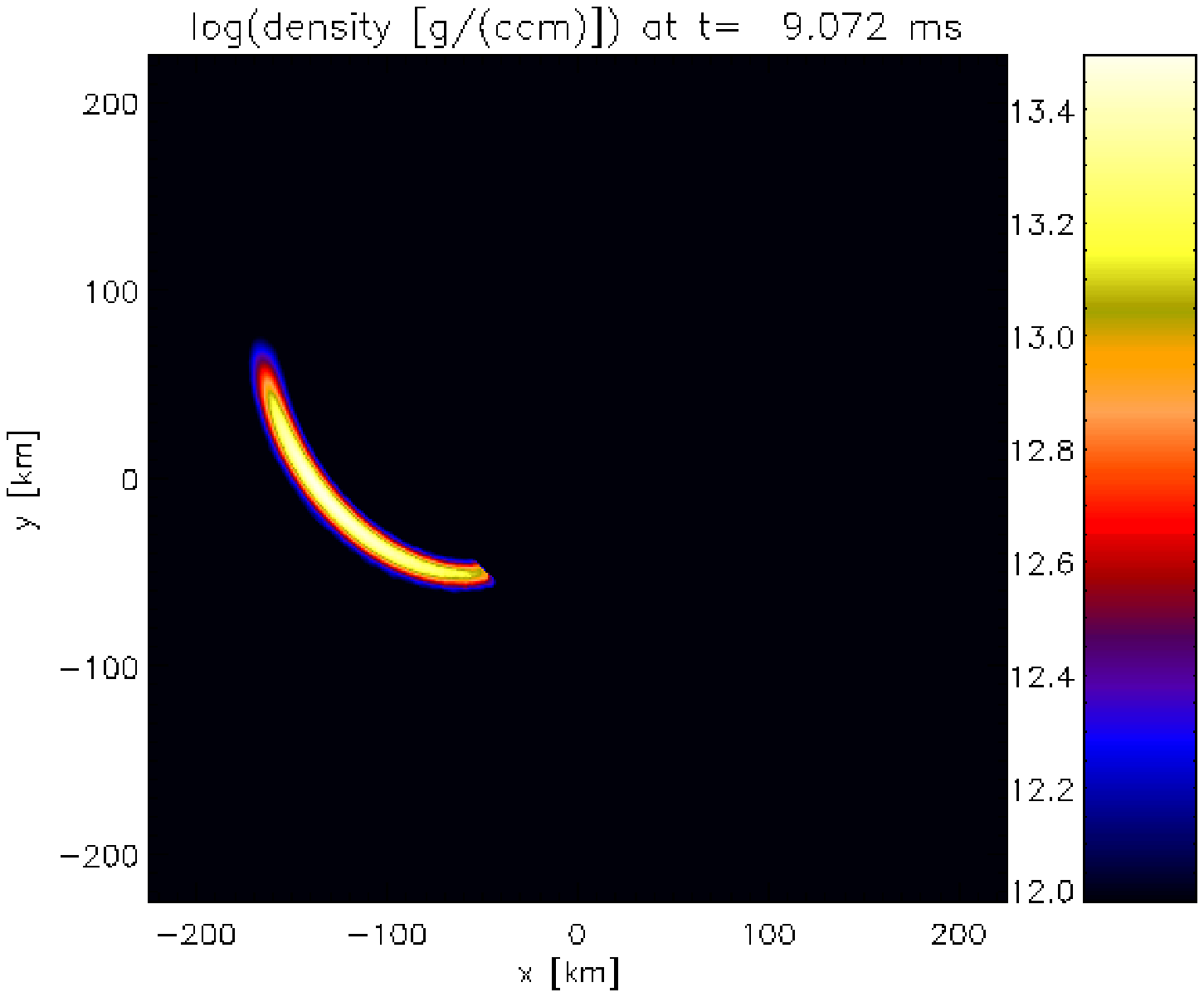,width=7.cm,angle=0}}
    \centerline{\psfig{file=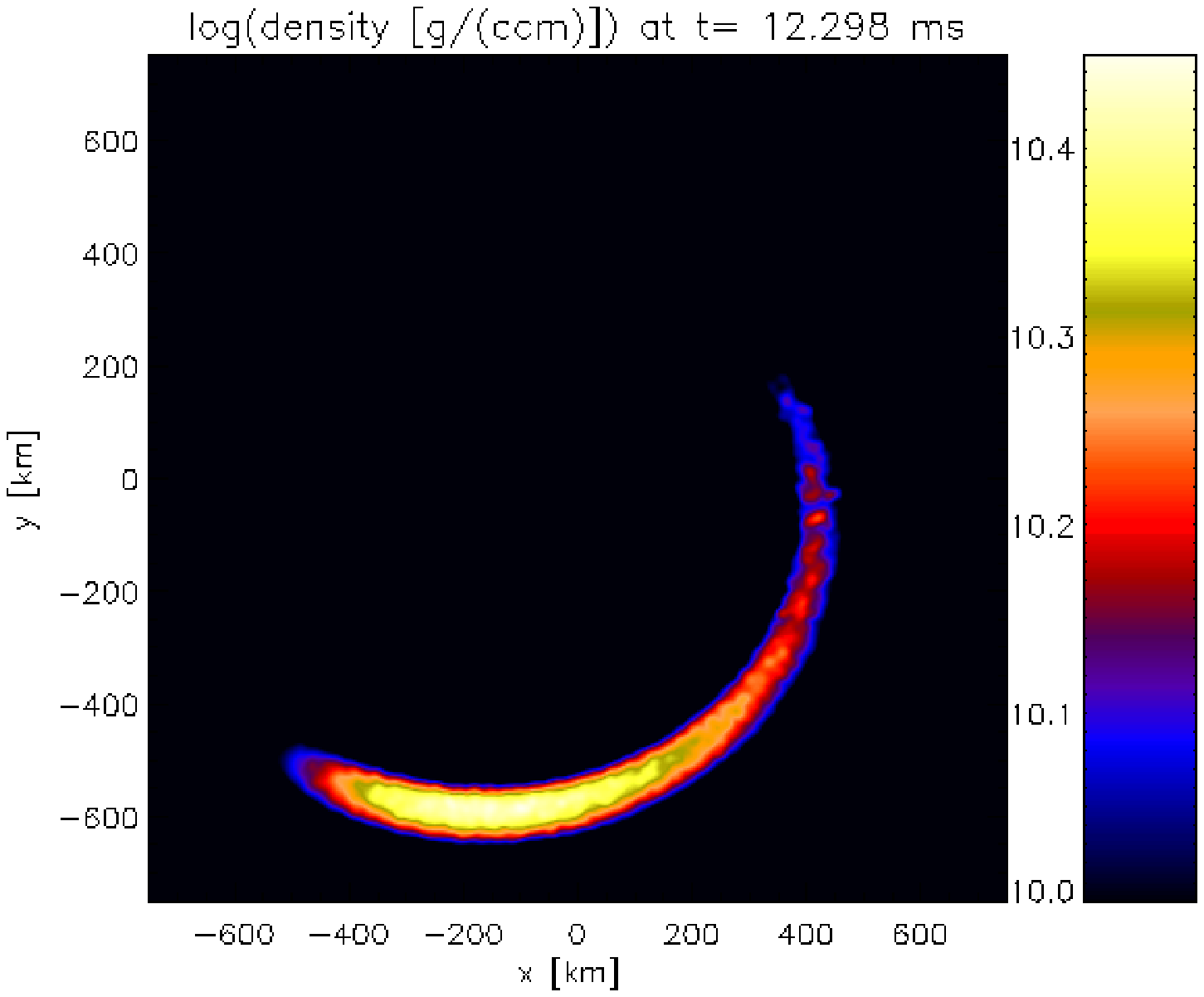,width=7.cm,angle=0}
                \psfig{file=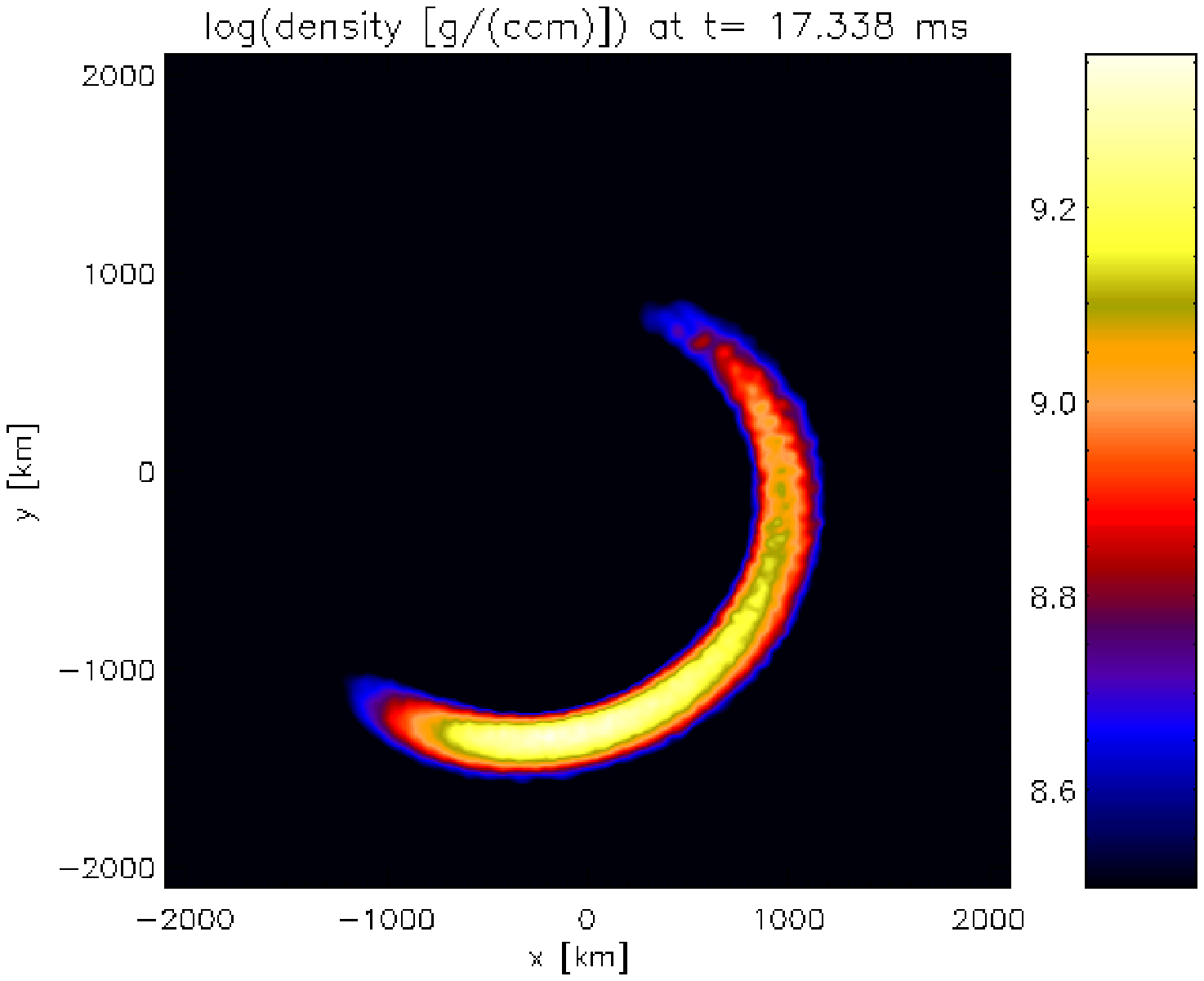,width=7.cm,angle=0}}
    \centerline{\psfig{file=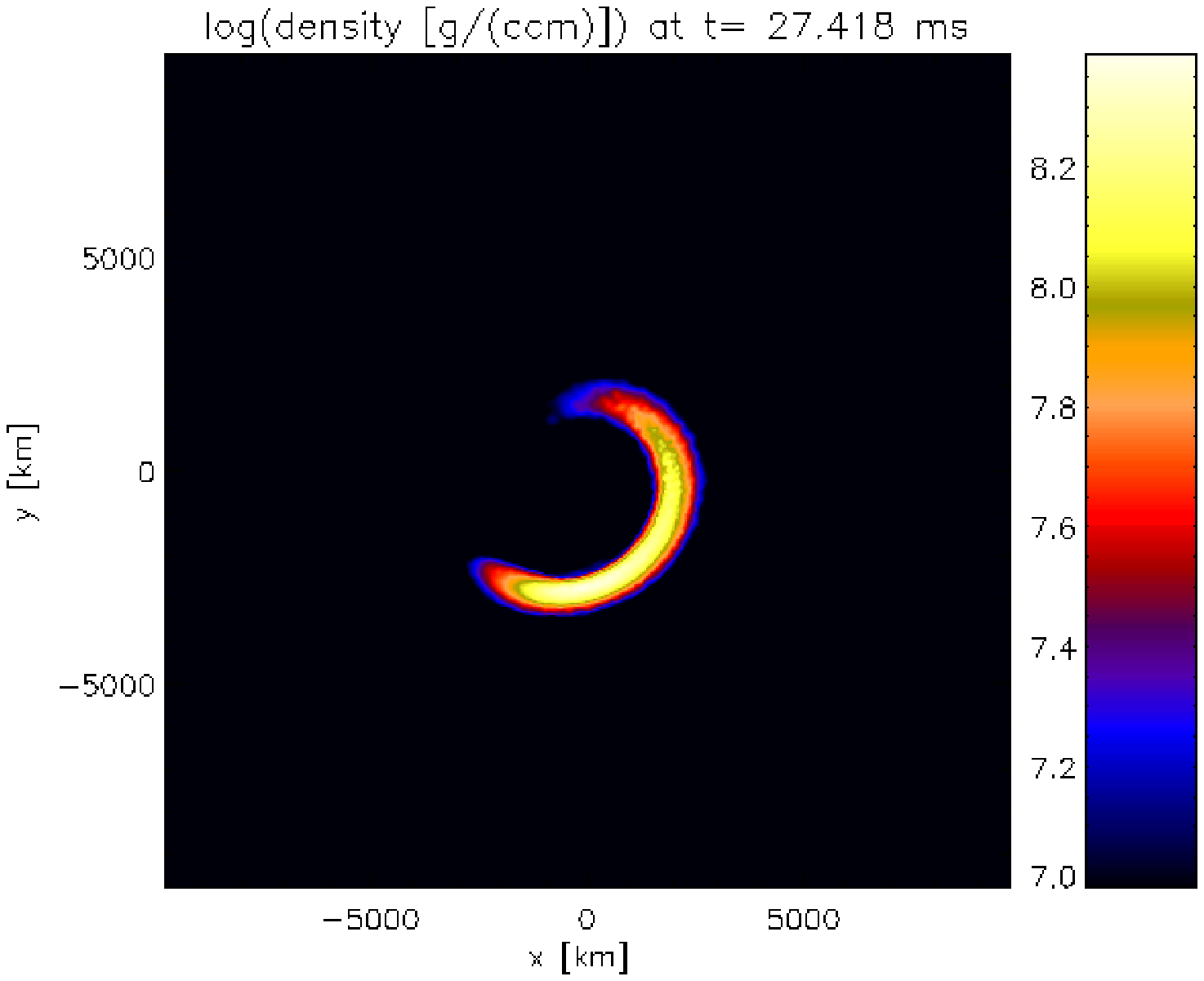,width=7.cm,angle=0}
                \psfig{file=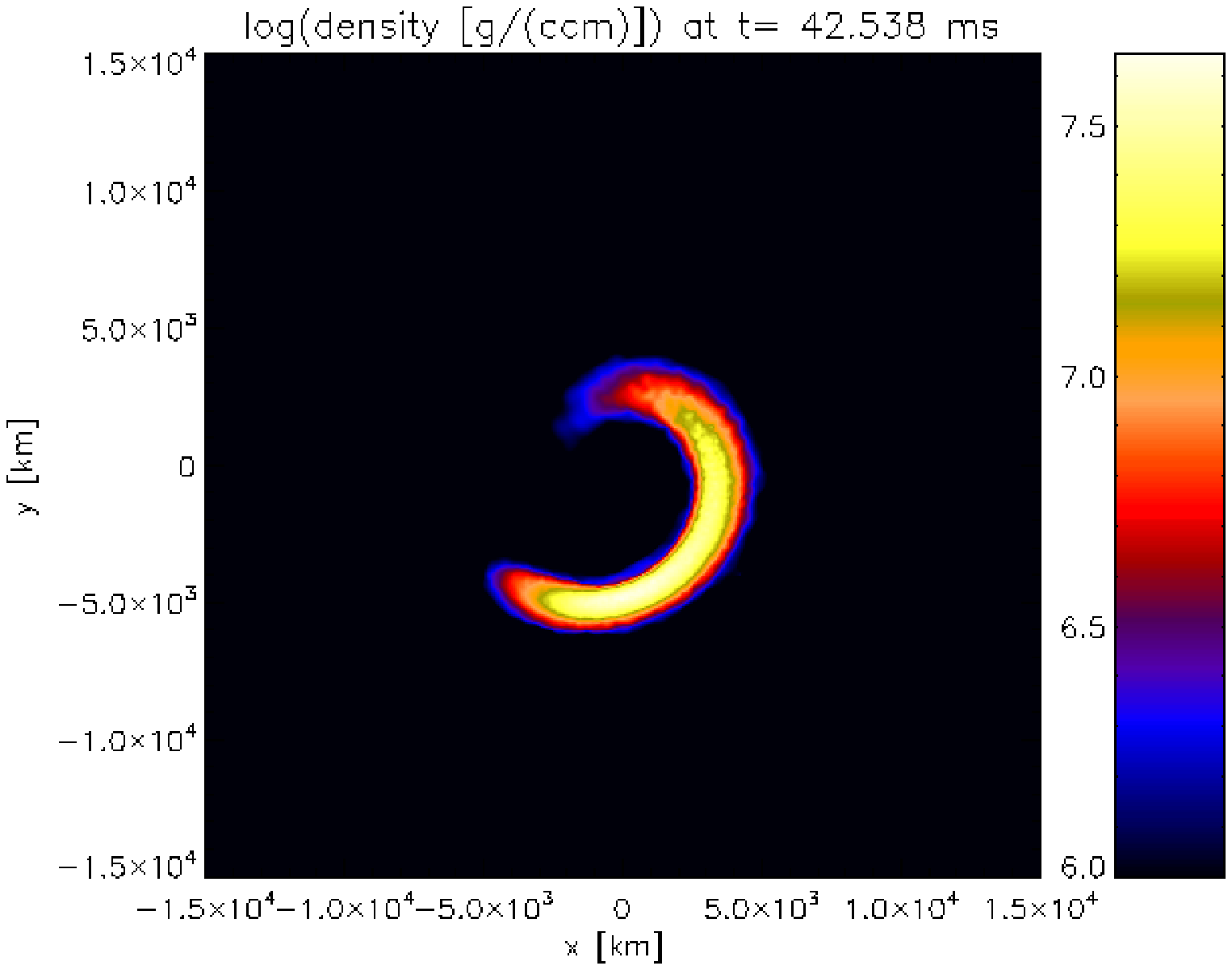,width=7.cm,angle=0}}
    \caption{\label{run VI}Density in the orbital plane of Run V (q=0.0778,
                tidally locked neutron star). }
\end{figure}

\clearpage

\begin{figure}
{\includegraphics[angle=-90,scale=.3]{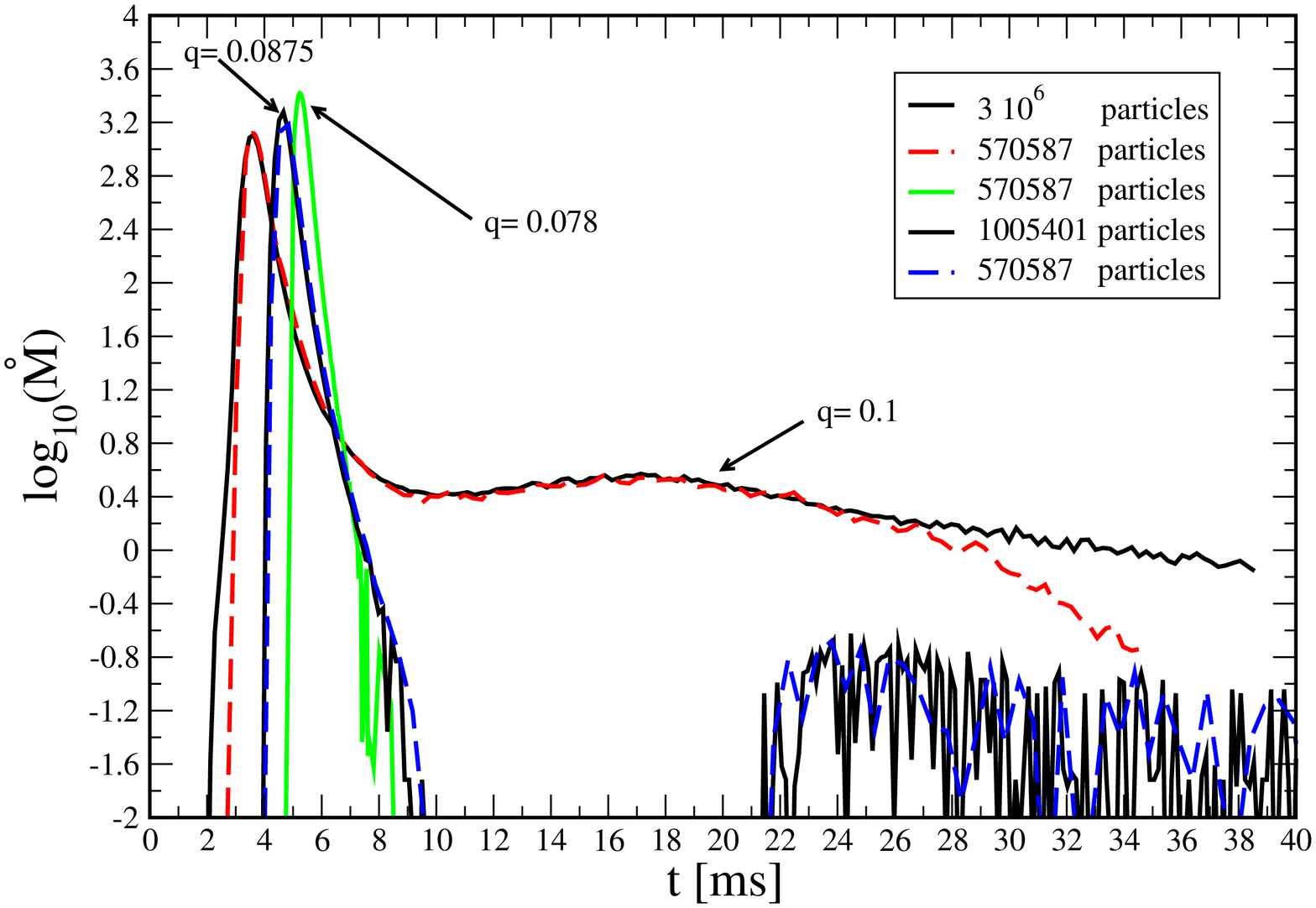}\includegraphics[angle=-90,scale=.3]{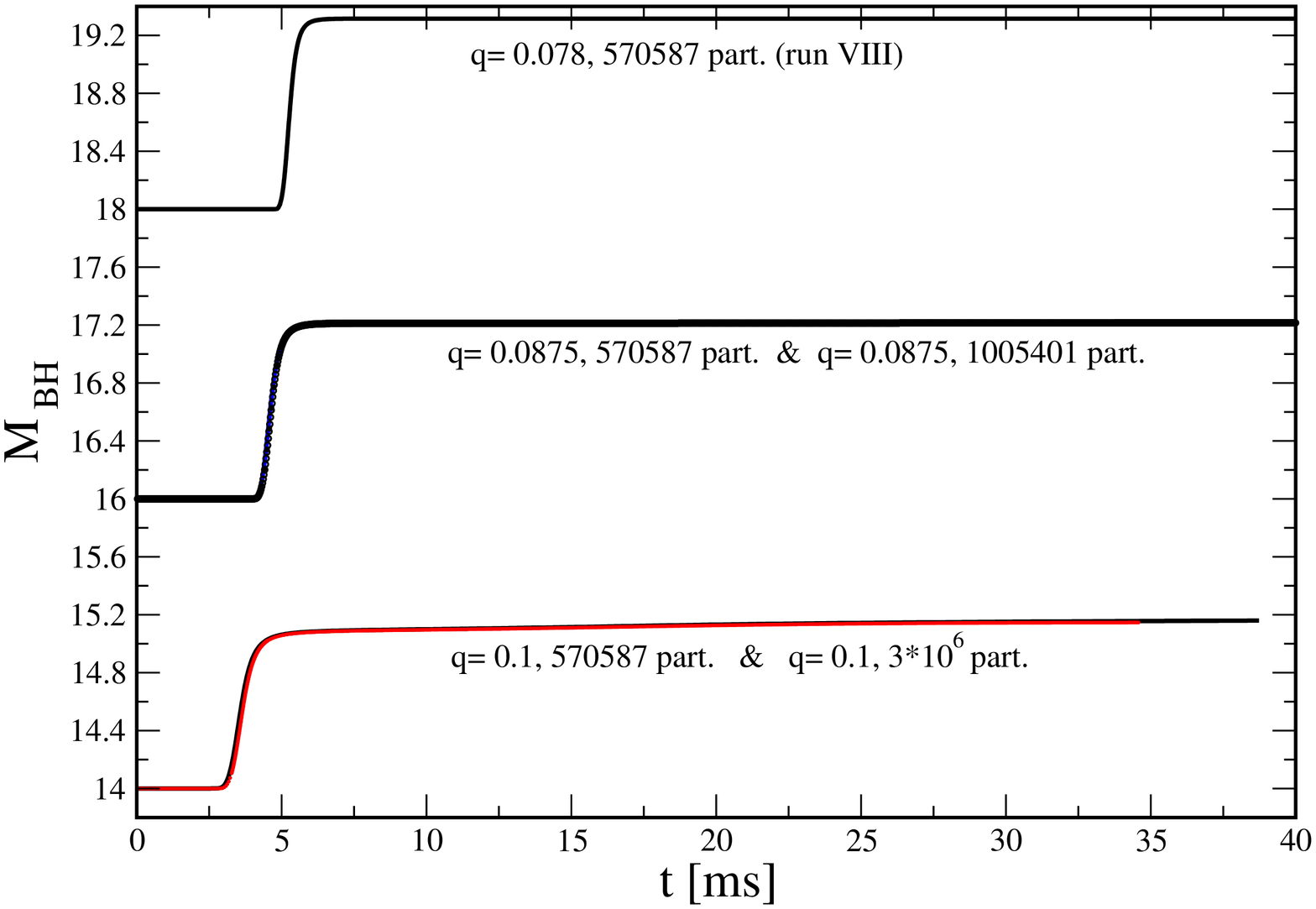}}
\caption{\label{mdot}
Left panel: The mass transfer rates as a function of time. In one case
(q=0.078, i.e M$_{\rm BH}$= 18 \msun) the mass transfer stops completely. This
is also true for the 20 \Msun case (not shown). 
Right panel: The growth of the black hole with time is shown for
  five of the runs.  Note that the runs that simulate the same
  systems (run I and II; run III and IV) with different resolutions yield
  nearly identical curves.} 
\end{figure}

\clearpage
\begin{figure}
\centerline{\psfig{file=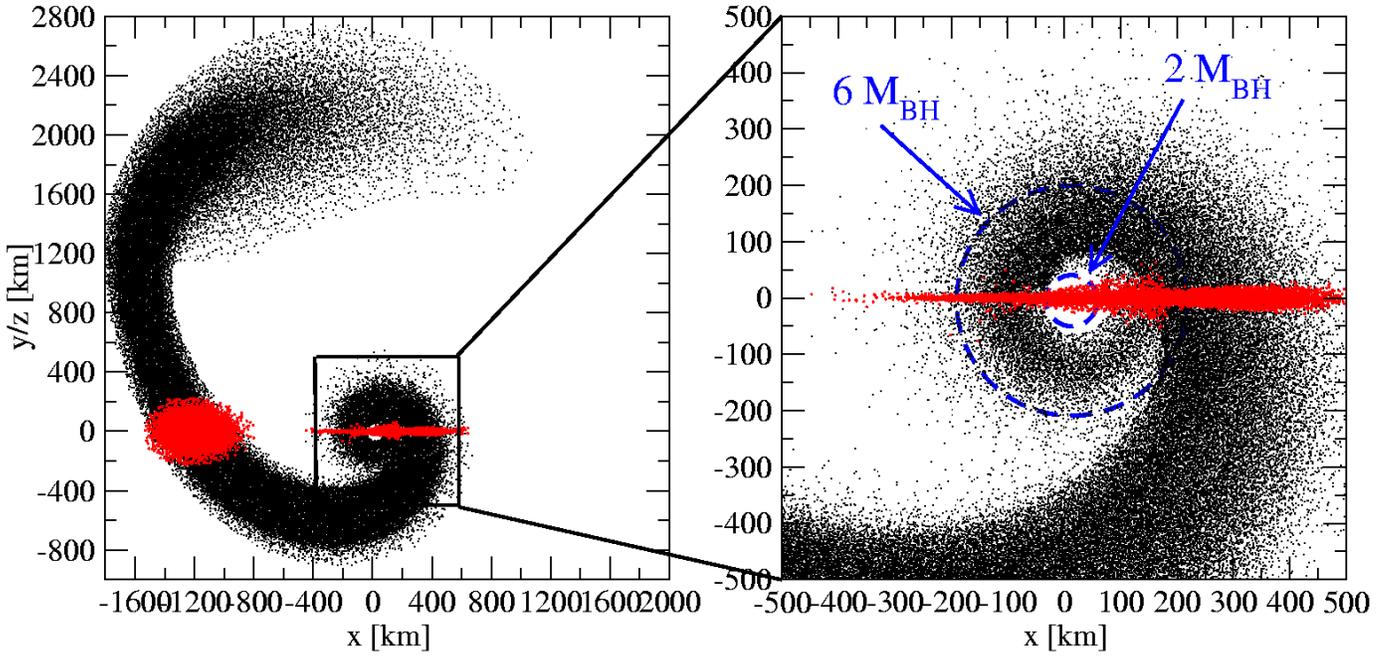,width=15cm,angle=-90}
 }
\caption{\label{particles_in_disk}Shown are the projections of the SPH
 particle positions onto the orbital plane (run II at t=18.396 ms). Overlaid
 are the projections to the X-Z-plane of those particles with $|y_i| < 150$
 km and the positions of the Schwarzschild-radius and the innermost stable
 circular orbit of a test particle around a Schwarzschild black hole. In
 regions of high column number densities only a fraction of the particles are
 displayed.}
\end{figure}            

\begin{figure}
\centerline{\psfig{file=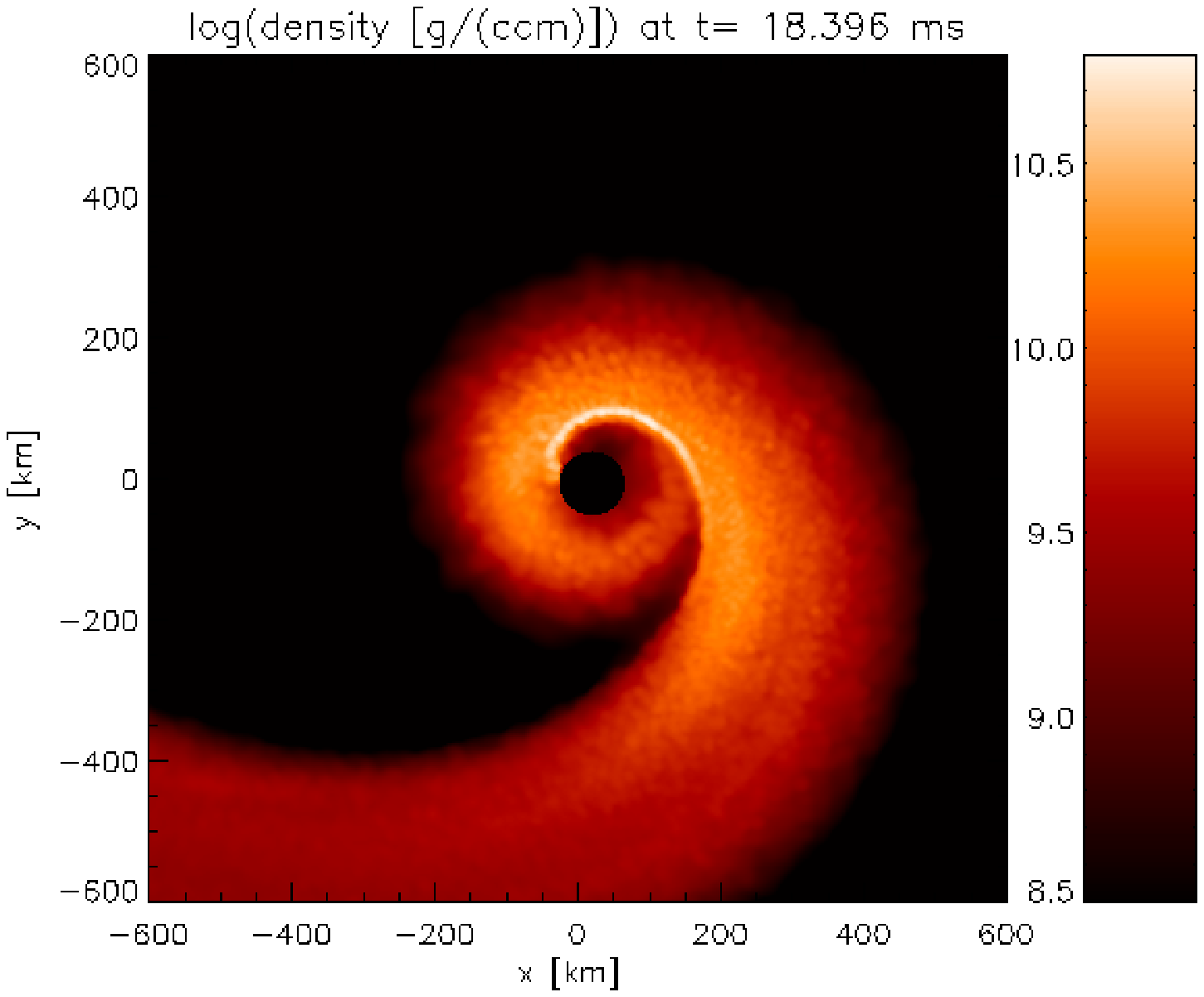,width=9cm,angle=0}\psfig{file=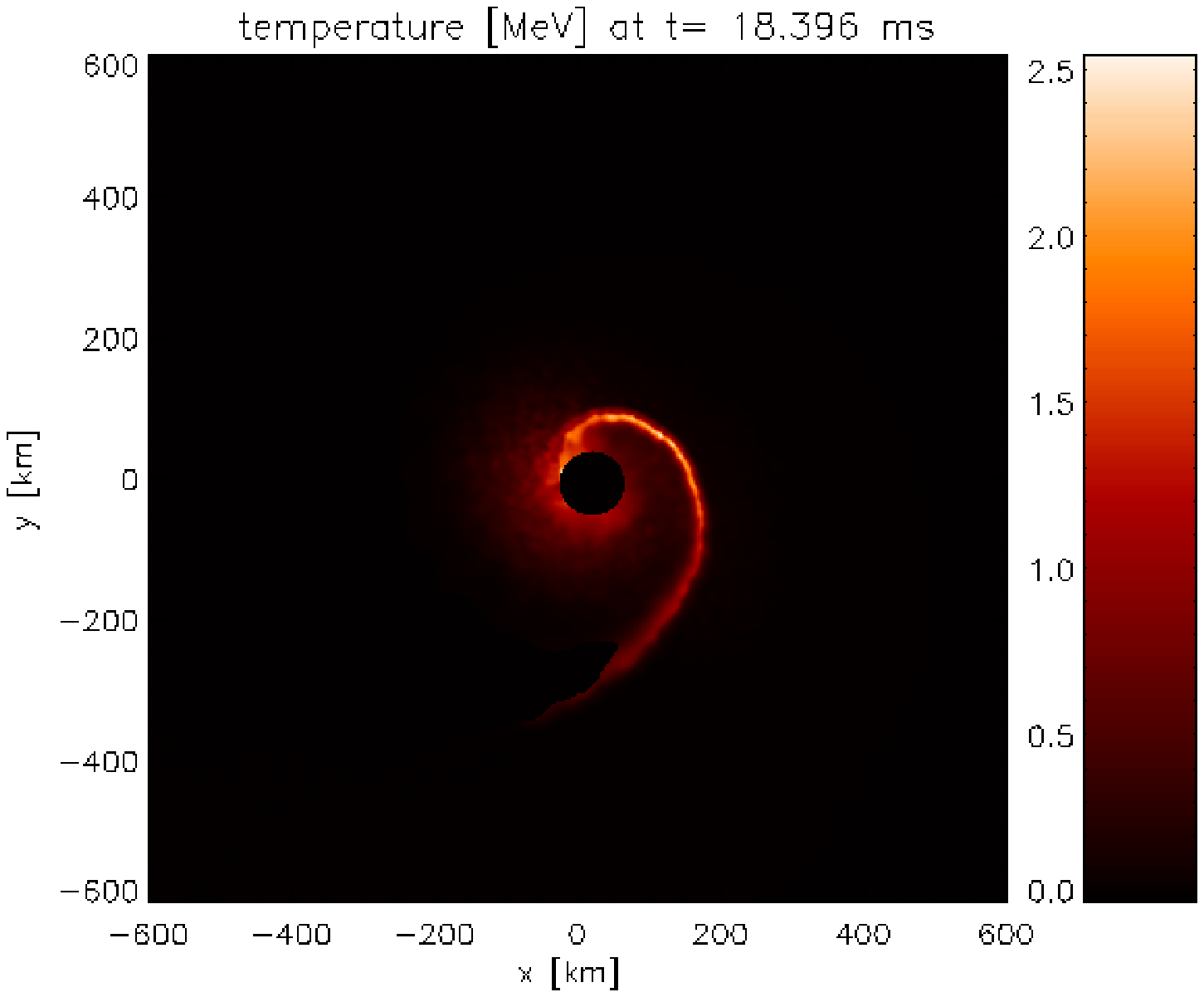,width=9cm,angle=0}
 }
\caption{\label{rho_T_runII}Blow-up of the inner disk region of run II at
  t=18.396 ms after simulation start (left panel: log(density); right panel: temperature). Clearly visible are the shock, where the
  accretion stream interacts with itself.}
\end{figure}

\begin{figure}
\centerline{\psfig{file=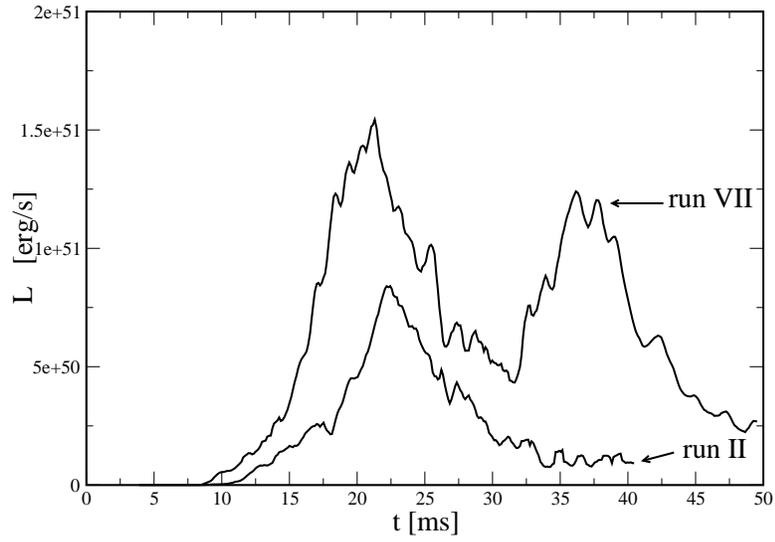,width=9cm,angle=-90}}
\caption{\label{Lnu_tot} Shown are the total neutrino luminosities of runs II
  (q= 0.1, tidal locking) and run VII (q= 0.1, no neutron star spin).}
\end{figure} 

\clearpage
\begin{figure}
\centerline{\psfig{file=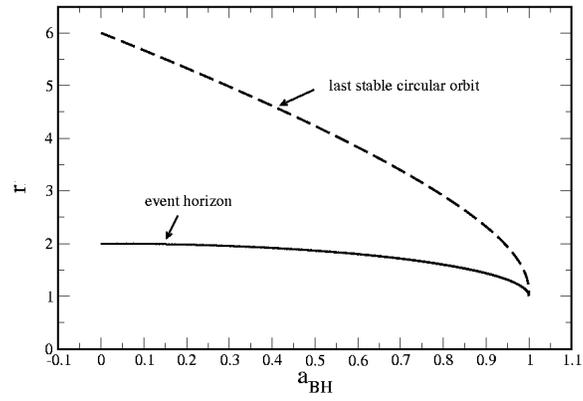,width=10cm,angle=0}}
\caption{\label{event_horizon}Position of the event horizon and the last
  stable circular orbit around a spinning black hole (see Novikov and Frolov
  1989) as a function of the spin parameter a$_{\rm BH}= J_{\rm BH}/M_{\rm
  BH}^2$.}
\end{figure} 

\clearpage

\begin{figure}
\centerline{\psfig{file=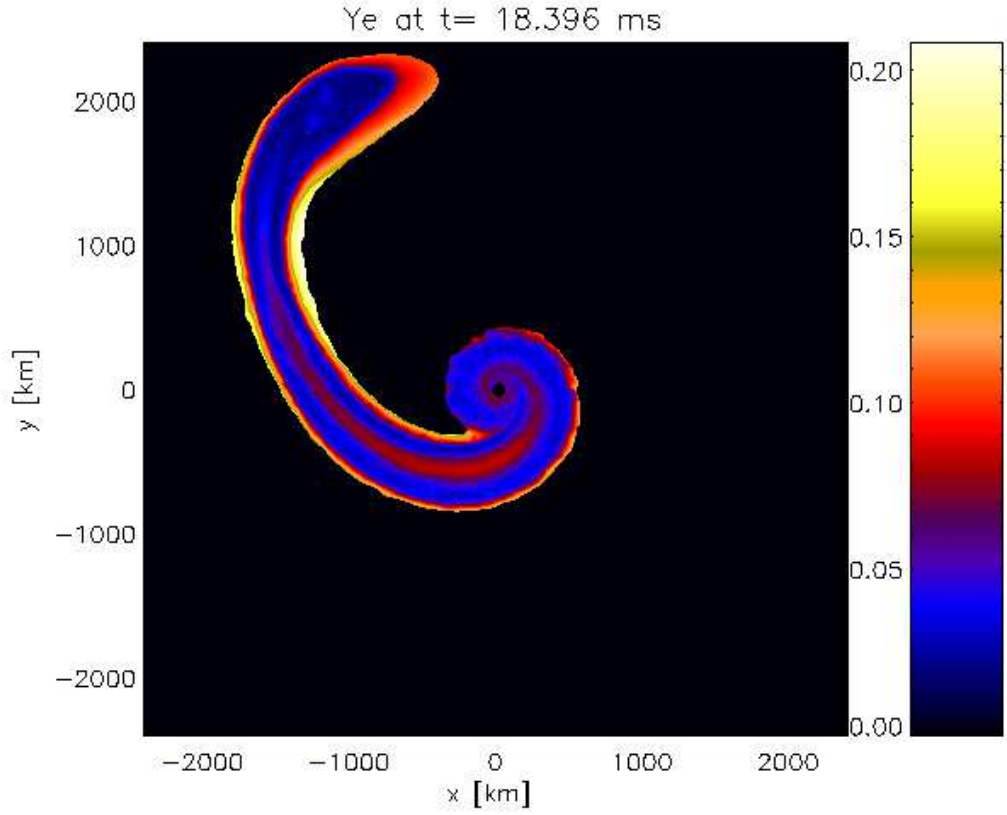,width=16cm,angle=0}
 }
\caption{\label{Ye_dist} Electron fraction, $Y_e$, in the orbital plane (run
 II, q= 0.1, tidal locking). The
 high-$Y_e$ skin around the remnant is the initial neutron star crust.}
\end{figure}  

\clearpage

\begin{figure}
\centerline{\psfig{file=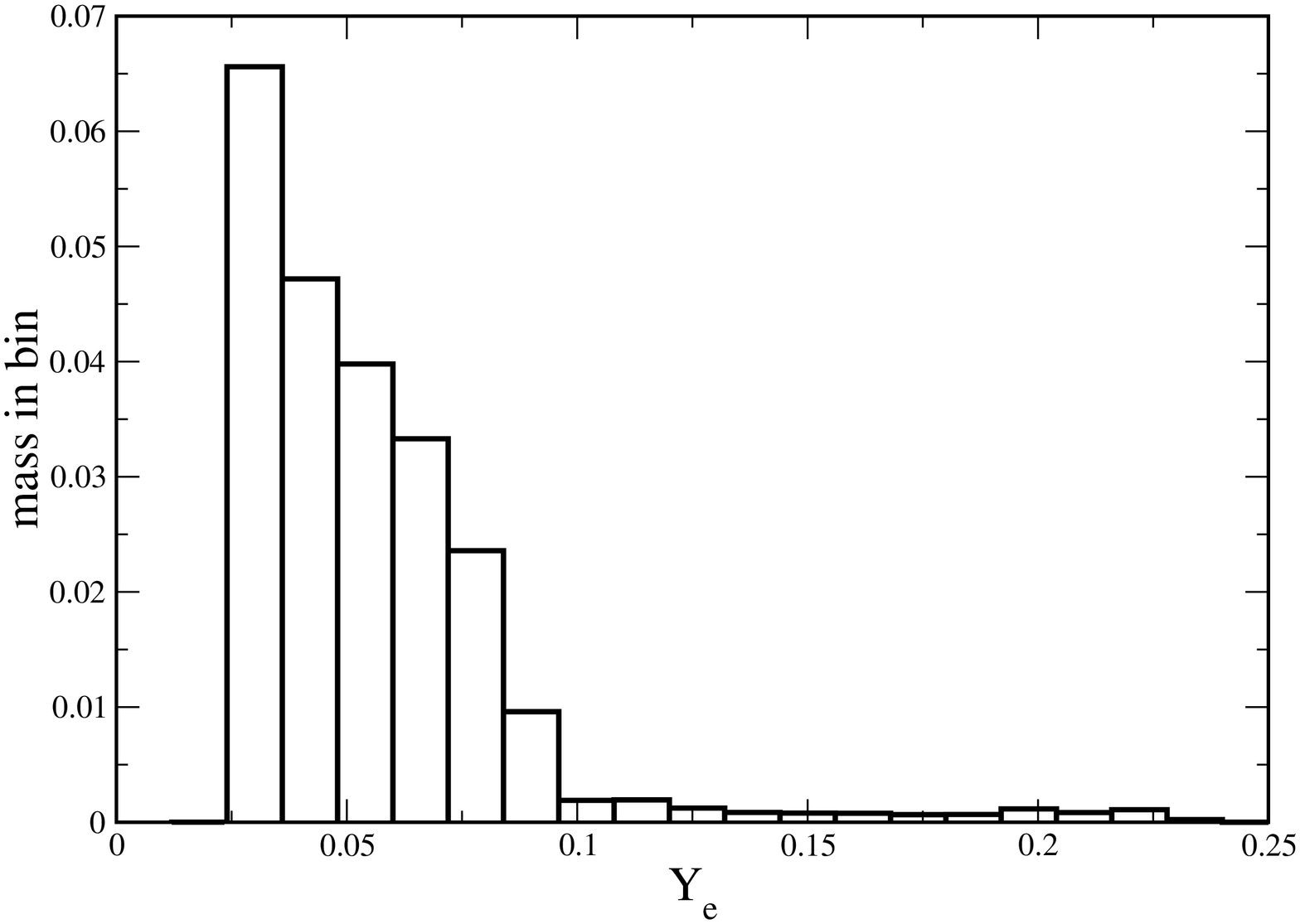,width=7cm,angle=-90}
            \psfig{file=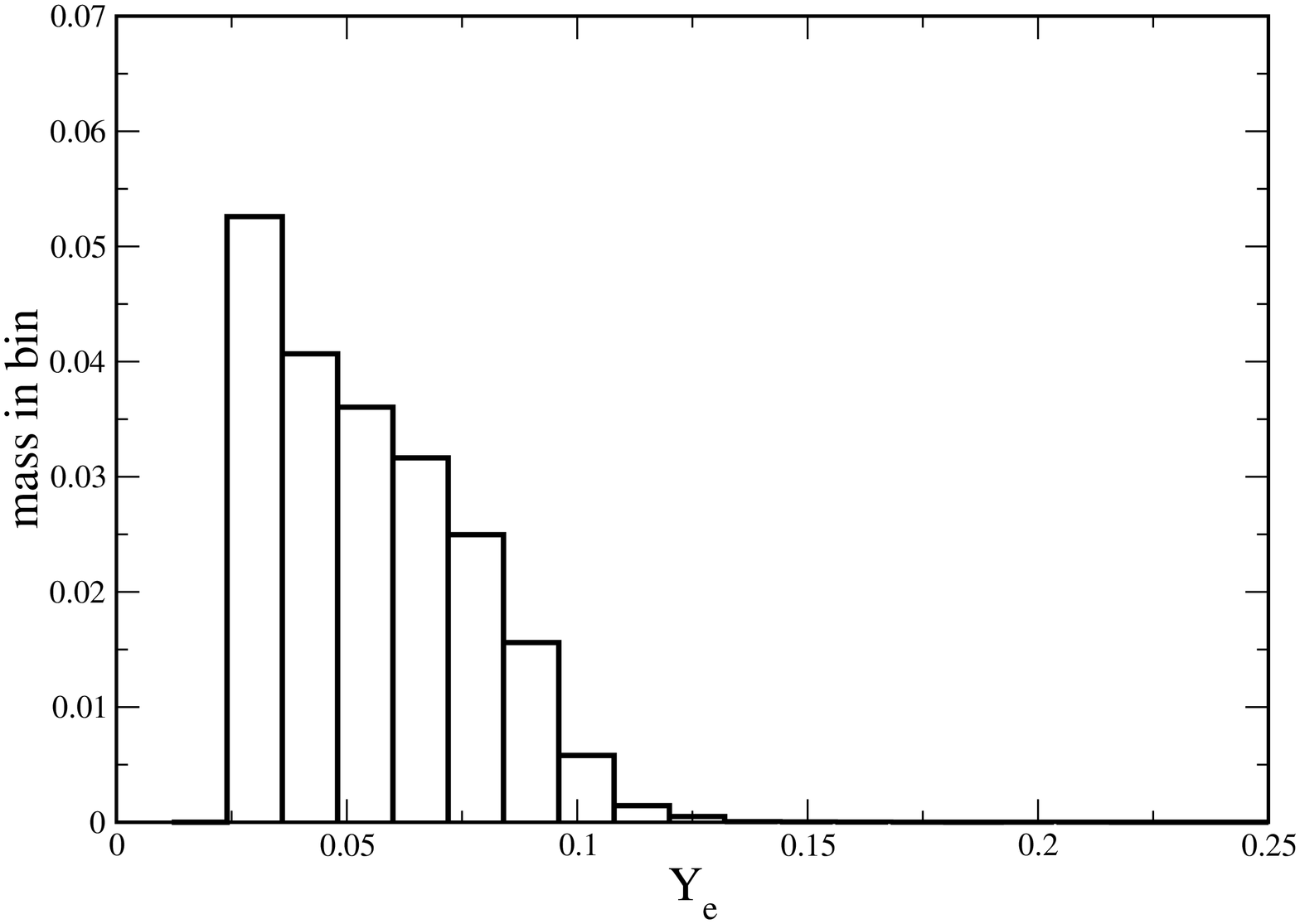,width=7cm,angle=-90}
 }
\caption{\label{Ye_binned}Histogram of the ejected mass binned according to
            $Y_e$. The left panel refers to run II (q= 0.1, tidal locking),
            the right one to run VII (q= 0.1, no neutron star spin). The
            missing high-$Y_e$ tail in the right panel may
            be an effect of the somewhat lower numerical resolution.} 
\end{figure} 

\clearpage

\begin{figure}
\centerline{\psfig{file=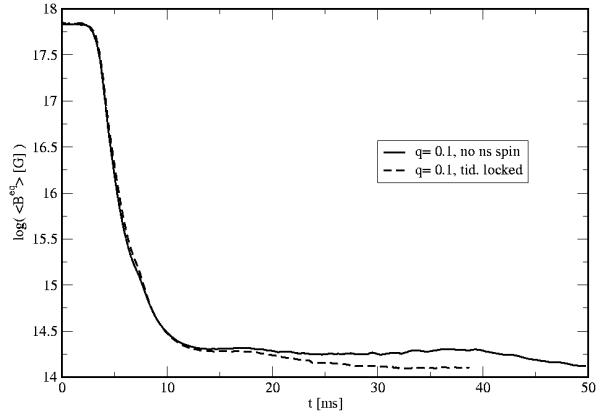,width=7cm,angle=-90}}
\caption{\label{Beq} Shown are the equipartition field strengths for run II
  (q= 0.1, tidal locking; dashed line) and run VII (q= 0.1, no neutron star
  spin; solid line) of the
  neutron star material averaged over the innermost 600 km (``disk'') as a
  function of time.} 
\end{figure}

\clearpage

 \begin{figure}
\centerline{\psfig{file=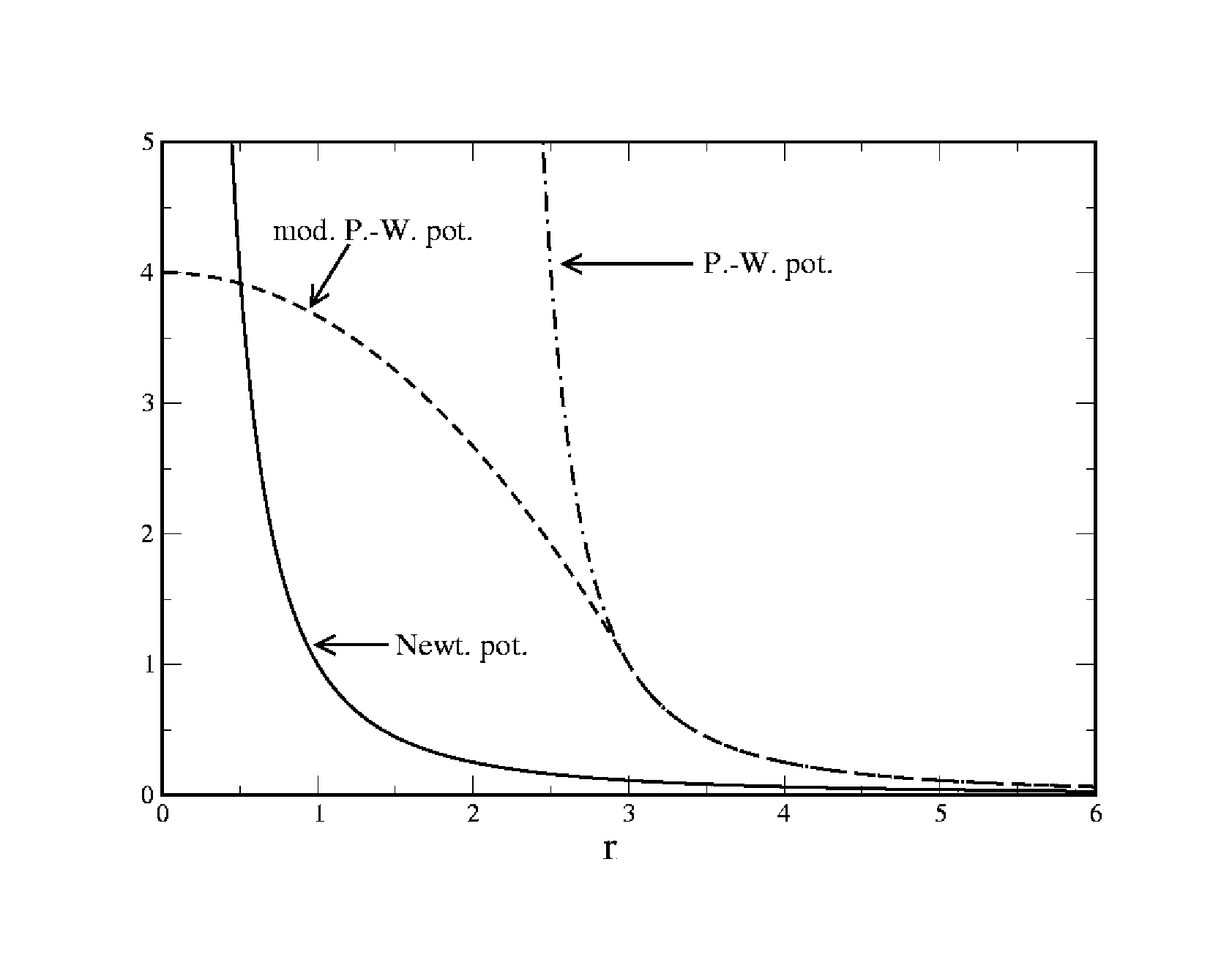,width=12cm,angle=0}}
\caption{Comparison of the denominator for purely Newtonian, \Pacz-Wiita
  and modified \Pacz-Wiita forces introduced to avoid the singularity at 2
  $M_{\rm BH}$ (for a transition radius, $R_t= 3 M_{\rm BH}$; see Appendix for
  details). All particles that have ever been inside $R_t$ are removed.}
\label{denom}
\end{figure}


\begin{thebibliography}{99}
\bibitem[Abramovici et al. 1992]{abramovici92}
Abramovici A.,  {Althouse} W.~E.,  {Drever} R. W.~P.,  {Gursel} Y.,  {Kawamura}
  S.,  {Raab} F.~J.,  {Shoemaker} D.,  {Sievers} L.,  {Spero} R.~E.,
  {Thorne} K.~S., Science, 256, 325 (1992)
\bibitem{abramowicz96}
Abramowicz, M.A., Beloborodov, A.M., Chen, X.M. and Ihumenshchev, I.V., A\&A,
313, 334 (1996)
\bibitem[Argast et al. 2004]{argast04}
Argast, D., Samland, M., Thielemann, F.-K. and Qian, Y..-Z., \aap, 416, 997
(2004) 
\bibitem{asano01}
 Asano, K. and Fukuyama, T., ApJ, 546, 1019 (2001)
\bibitem{balbus98}
Balbus, S.A. and Hawley, J.F., Rev. Mod. Phys., 70, 1 (1998)
\bibitem{balsara95}
Balsara D., J. Comput. Phys., 121, 357 (1995)
\bibitem{benz90b}
Benz W.,  Bowers R.,  Cameron A.,    Press W.,  1990, ApJ, 348, 647
\bibitem{bloom05}
Bloom, J.S. et al., astro-ph/0505480 (2005)
\bibitem{chandra60}
Chandrasekhar, S., Proc. Nat. Acad. Sci USA, 46, 53 (1960)
\bibitem{fehlberg68}
Fehlberg, E., NASA Technical Report TR-R287 (1968)
\bibitem{janka99}
Janka, H.-T., Eberl, T., Ruffert, M., Fryer, C.L., ApJ, 527, L39 (1999)
\bibitem{lattimer74}
Lattimer, J. and Schramm, D.N., ApJ, 192, L145 (1974)
\bibitem{lattimer76}
Lattimer, J. and Schramm, D.N., ApJ, 210, L549 (1976)
\bibitem{lattimer77}
Lattimer, J. et al., ApJ, 213, L225 (1977)
\bibitem{lee99a}
Lee, W.H. and Kluzniak, W.L., ApJ, 526, 178 (1999)
\bibitem{lee99b}
Lee, W.H. and Kluzniak, W.L., MNRAS, 308, 780 (1999)
\bibitem{lee00}
Lee, W.H., MNRAS, 318, L606 (2000)
\bibitem{lee01}
Lee, W.H., MNRAS, 328, 583 (2001)
\bibitem{lee05a}
Lee, W.H., Ramirez-Ruiz, E. and Page, D., astro-ph/0506121 (2005)
\bibitem{lee05b}
Lee, W.H., Ramirez-Ruiz, E. and Granot, J., astro-ph/0506104 (2005)
\bibitem{miller05}
Miller, M.C., astro-ph/0505094 (2005)
\bibitem{morris97}
Morris J.,  Monaghan J., J. Comp. Phys., 136, 41 (1997)
\bibitem{novikov89}
Novikov, I.D. and Frolov, V.P., Physics of Black Holes, Dordrecht: Kluwer
(1989) 
\bibitem{perna02}
Perna, R. and  Belczynski, K., ApJ, 570, 252 (2002)
\bibitem{pfahl05}
Pfahl, E., Podsiadlowski, P. and Rappaport, S., astro-ph/0502122
\bibitem{piran99}
Piran, T., Phys. Rep., 314, 575 (1999)
\bibitem{ramirez-ruiz05}
Ramirez-Ruiz, E. and Socrates, A., astro-ph/0504/257 (2005)
\bibitem{rasio05}
Rasio, F., Faber, J., Kobayashi, S. and Laguna, P., astro-ph/0503007 (2005)
\bibitem{rosswog00}
Rosswog S.,  Davies M.~B.,  Thielemann F.-K.,    Piran T.,  A \&\ A, 360,
  171 (2000)
\bibitem{rosswog02a}
Rosswog, S. and Davies, M.-B., MNRAS, 334, 481 (2002)
\bibitem{rosswog2002b}
Rosswog, S. and Ramirez-Ruiz, E. MNRAS, 336, L7 (2002)
\bibitem{rosswog03a}
Rosswog, S. and Liebend\"orfer, M., MNRAS, 342, 673 (2003)
\bibitem {rosswog03b}
Rosswog, S., Ramirez-Ruiz, E. and Davies, M.B., MNRAS 345, 1077 (2003)
\bibitem{rosswog03c}
Rosswog, S. and Ramirez-Ruiz, E., MNRAS, 343, L36 (2003)
\bibitem{setiawan04}
Setiawan, S., Ruffert, M., Janka, H.-T., MNRAS, 352, 753 (2004)
\bibitem{shen98a}
Shen H.,  Toki H.,  Oyamatsu K.,    Sumiyoshi K., Nuclear Physics, A
  637, 435 (1998)
\bibitem{shen98b}
Shen H., Toki H.,  Oyamatsu K., Sumiyoshi K., Prog. Theor. Phys., 100, 1013 
(1998)
\bibitem{stairs04}
Stairs, I.H., Science, 304, 547 (2004)
\bibitem{Taniguchi05}
Taniguchi, K., Baumgarte, T.W., Faber, J.A. and Shapiro, S.L.,
astro-ph/0505450v2 (2005)
\bibitem{velikhov59}
Velikhov, E.P., Sov. Phys. JETP., 36, 995 (1959)
\end{thebibliography}
\end{document}